\documentclass[apj,twocolumn]{emulateapj}
\usepackage{natbib,amssymb,amsmath,graphicx}
\shorttitle{Inverse Compton Scattering and Coronal HXR Sources}
\shortauthors{Chen \& Bastian}

\begin{document}

\title{The Role of Inverse Compton Scattering in Solar Coronal \\ 
Hard X-ray and $\gamma$-ray Sources}

\author{Bin Chen}
\affil{Department of Astronomy, University of Virginia, Charlottesville, VA 22904 USA}

\author{T. S.  Bastian}
\affil{National Radio Astronomy Observatory, Charlottesville, VA 22903}

\begin{abstract}
\noindent Coronal hard X-ray (HXR) and continuum $\gamma$-ray sources associated with the impulsive phase of solar flares have been the subject of renewed interest in recent years. They have been interpreted in terms of thin-target, nonthermal bremsstrahlung emission. This interpretation has led to rather extreme physical requirements in some cases. For example, in one case, essentially all of the electrons in the source must be accelerated to nonthermal energies to account for the coronal HXR source. In other cases, the extremely hard photon spectra of the coronal continuum $\gamma$-ray emission suggest that the low energy cutoff of the electron energy distribution lies in the MeV energy range. Here we consider the role of inverse Compton scattering (ICS) as an alternate emission mechanism in both the ultra- and mildly relativistic regimes. It is known that relativistic electrons are produced during powerful flares; these are capable of up-scattering soft photospheric photons to HXR and $\gamma$-ray energies. Previously overlooked is the fact that mildly relativistic electrons, generally produced in much greater numbers in flares of all sizes, can up-scatter EUV/SXR photons to HXR energies. We also explore ICS on anisotropic electron distributions and show that the resulting emission can be significantly enhanced over an isotropic electron distribution for favorable viewing geometries. We briefly review results from bremsstrahlung emission and reconsider circumstances under which nonthermal bremsstrahlung or ICS would be favored. Finally, we consider a selection of coronal HXR and $\gamma$-ray events and find that in some cases the ICS is a viable alternative emission mechanism.  
\end{abstract}


\section{Introduction}

Solar flares produce HXR and continuum $\gamma$-ray emission, generally attributed to thermal or  non-thermal bremsstrahlung emission. As such, HXR emission provides key diagnostics of plasma heating, electron acceleration, and electron transport.  Intense chromospheric HXR thick target emission is produced at the foot points of coronal magnetic loops. The relatively faint coronal HXR and continuum $\gamma$-ray emission is generally more difficult to observe in the presence of intense foot point emission given the limited dynamic range of X-ray imaging instruments. In most cases, therefore, coronal HXR or $\gamma$-ray emission is observed in flares that occur in active regions behind the solar limb; the intense foot point emission is occulted, thereby revealing the relatively faint coronal emission. While reports of coronal HXR emission data back to the early 1970s \citep[e.g.,][]{1971ApJ...165..655F} imaging observations, first by {\sl Yohkoh} \citep{1992PASJ...44L..45K} and then by the {\sl Ramaty High-Energy Solar Spectroscopic Imager} \citep[RHESSI; ][]{2002SoPh..210....3L}, have led to renewed interest in coronal HXR sources. 

As discussed in the review by  \citet{2008A&ARv..16..155K}, coronal HXR sources reveal a diverse phenomenology. These include sources that precede the impulsive phase \citep{2003ApJ...595L..69L} as well as a variety of coronal HXR sources that may occur during the impulsive phase: ``over-the-loop-top" sources \citep{1994Natur.371..495M}, double sources \citep{2003ApJ...596L.251S}, and coronal thick target sources \citep{2004ApJ...603L.117V}. During the late phase of flares ``superhot" thermal sources \citep{1981ApJ...251L.109L}, gradual sources that display a ``soft-hard-harder" spectral evolution \citep{1986ApJ...305..920C,1995ApJ...453..973K}, and non-thermal sources that display hard, continuum $\gamma$-ray emission \citep{2008ApJ...678L..63K} may occur. Coronal HXR sources have been observed over a range of heights in the corona and can be associated with stationary or moving sources. We refer the reader to \citet{2008A&ARv..16..155K}, and references therein, for a more detailed discussion of the types of coronal HXR sources, their properties, and the circumstances under which they occur. 

We focus here on non-thermal  coronal HXR and continuum $\gamma$-ray sources that occur during the impulsive phase of flares.  These have been interpreted in terms of thin-target, non-thermal bremsstrahlung. This may well be correct but in the case of certain coronal HXR, or continuum $\gamma$-ray, sources the parameters required can be extreme. For example, three powerful X-class flares - those on 2003 Oct 28, 2005 Jan 20, and 2005 Sep 7 - were accompanied by continuum $\gamma$-ray emission $>200$ keV \citep{2008ApJ...678L..63K}. These flares, observed by RHESSI, were not occulted by the limb and both foot point and coronal emissions were observed. In the case of 2005 Jan 20 foot point emission dominated during times near the $\gamma$-ray maximum but the coronal source became increasingly prominent during the decay of the $\gamma$-ray emission. The power-law index $\alpha$ of the photon spectrum of the coronal source from 200-800 keV was significantly harder ($\alpha\approx 1.5$) than that of the foot points ($\alpha\approx 2.9$). The other two flares displayed similar properties. Interpreting the emission in terms of non-thermal, thin-target, electron-ion bremsstrahlung emission implies the spectral indices of the coronal $\gamma$-ray sources were near the minimum values theoretically possible and require the effective low-energy cutoff of the energetic electrons responsible for the emission to have been $>1$ MeV in all cases \citep{2008A&A...486.1023B}. Another intriguing example is the observation of an HXR source high in the corona \citep{2007ApJ...669L..49K}. The flare itself occurred $40^\circ$ behind the limb; to be visible from the Earth the HXR source was at a radial height of order 150~Mm where the ambient density was estimated to be only $\sim 10^8$ cm$^{-3}$. In order to produce the diffuse HXR source via thin-target bremsstrahlung roughly 10\% of the electrons were accelerated. Yet more extreme is an event recently reported by \citet{2010ApJ...714.1108K} that bears a resemblance to the celebrated ``Masuda flare'' \citep{1994Natur.371..495M}. Also observed by RHESSI, the limb-occulted flare on 2007 Dec 31 showed a relatively intense nonthermal coronal HXR source located $\approx 6$ Mm above the thermal flare loops. The HXR source can be understood in terms of non-thermal, thin-target bremsstrahlung emission from a power-law distribution of electrons if essentially all of the electrons in the source are accelerated to nonthermal energies. It is worth asking whether an alternate emission mechanism is responsible for, or may contribute to, some coronal HXR sources. 

In considering plausible mechanisms for HXR emission from solar flares, \citet{1971SoPh...18..284K} considered non-thermal bremsstrahlung emission, synchrotron emission, and inverse Compton scattering (ICS). He concluded that bremsstrahlung emission is favored in most cases but that ICS could play a role under some circumstances - if the ambient density in the source is low, for example. Synchrotron radiation is not favored as a mechanism for HXR emission from flares. In light of the many recent observations of coronal HXR sources, however, it seems timely to revisit the question of whether ICS plays a role. We are not the first to do so. Motivated in part by the observations cited above, \citet{2010A&A...510A..29M} recently considered whether photospheric photons up-scattered to HXR or $\gamma$-ray energies by relativistic electrons or positrons by ICS could account for coronal HXR sources. They consider scattering of the (anisotropic) photospheric photon field on an isotropic, power-law distribution of electrons from sources at various heliographic longitudes. They find that conditions for ICS are most favorable for coronal sources near the limb provided the ambient density is sufficiently low. They also find that relatively modest numbers of energetic electrons are required for ICS to account for the continuum $\gamma$-ray sources discussed by \citet{2008ApJ...678L..63K}. Unfortunately, the work contained an error that renders their estimates of the number density of energetic electrons required too optimistic. 

In this paper we correct and expand upon the work of \citet[][hereafter MM10]{2010A&A...510A..29M}. Unlike MM10, we do not consider ICS on positrons. We begin by considering, as they do, the case of ICS from an isotropic, power-law distribution of ultra-relativistic electrons scattering an anisotropic field of photons in \S2.1. Noting that flares produce copious extreme ultraviolet (EUV) and soft X-ray (SXR) emission, we next consider the case where EUV/SXR photons are up-scattered  to HXR or $\gamma$-ray energies by a power-law distribution of mildly relativistic electrons (\S2.2). To do so requires evaluating exact expressions for the ICS photon scattering rate. These calculations show that the ICS spectrum resulting from scattering on mildly relativistic electrons has steeper photon spectrum than that resulting from scattering on ultra-relativistic electrons. Finally, we note that anisotropic electron distributions may have significant implications for ICS. We consider ICS on ``beam" and ``pancake" electron distributions in \S3, showing that they can result in significant enhancements compared with ICS on isotropic electrons with the same number density for favorable viewing geometries. We briefly review results from anisotropic electron-ion and electron-electron bremsstrahlung in \S4 and reconsider circumstances under which nonthermal bremsstrahlung or ICS is favored. We discuss our results in light of selected observations of coronal HXR/$\gamma$-ray sources in \S5 and conclude in \S6.

\section{ICS of Photons on an Isotropic Electron Distribution}

In this section we consider ICS of isotropic and anisotropic photon distributions on isotropic distributions of ultra-relativistic electrons (Lorentz factor $\gamma\gg 1$) and mildly relativistic electrons ($\gamma\sim 2-$10). The vast majority of studies that consider ICS do so for astrophysical regimes where the electrons in question are ultra-relativistic and the ambient plasma density is very low.  In the ultra-relativistic case, certain approximations can be made that greatly simplify the relevant expressions for the photon distribution function or emissivity. We therefore begin by considering ultra-relativistic electrons interacting with a photon field. Two cases are considered: that in which the photons are taken to be isotropic and that in which they are taken to be anisotropic. We then consider the corresponding cases for mildly relativistic electrons. We take $\hbar=m_e=c=1$ throughout the paper. 

\subsection{Ultra-Relativistic Regime}

\citet{1968PhRv..167.1159J} first derived exact expressions for ICS from an isotropic distribution of ultra-relativistic monoenergetic electrons interacting with an isotropic distribution of monoenergetic photons normalized to one electron passing through a photon field of unit number density. The net rate at which photons are scattered into a particular energy $\epsilon_2$ is given by 
\begin{equation}
R_{iso}(\epsilon_2)=\int d\gamma n_e(\gamma)\int d\epsilon_1 n_\gamma(\epsilon_1) \frac{dR_{iso}(\gamma,\epsilon_1)}{d\epsilon_2}
\end{equation}

\noindent where $n_e(\gamma)$ is the electron number density distribution, $n_\gamma(\epsilon_1)$ is the number density distribution of the incident photons, and $dR_{iso}(\epsilon_1)/d\epsilon_2$ is the rate at which photons are scattered from $\epsilon_1$ to $\epsilon_2$ (note that $\epsilon_2$ may be greater than or less than $\epsilon_1$, in general). The time dependence is implicit. 

The exact expressions for the differential scattering rate are rather cumbersome although they have been presented in simplified form by \citet{2005ApJ...628..857P}. Jones shows that the exact expressions for the differential rates at which photons are up-scattered or down-scattered can be approximated by the much simpler expressions:

\begin{equation}\label{eq:jones}
\begin{split}
\frac{dR_{\rm iso}(\gamma,\epsilon_1)}{d\epsilon_2}= & \frac{2\pi r_e^2}{\epsilon_1\gamma^2}\left[2q\ln q+(1+2q)(1-q)\right.\\
&+\left.\frac{1}{2}\frac{(4\epsilon_1\gamma q)^2}{(1+4\epsilon_1\gamma q)}(1-q) \right] \quad (\epsilon_2>\epsilon_1);
\end{split}
\end{equation}   

\begin{equation}
\begin{split}
\frac{dR_{\rm iso}(\gamma,\epsilon_1)}{d\epsilon_2}= & \frac{\pi r_e^2}{2\epsilon_1\gamma^4}\left[(q'-1)(1+2/q')-2\ln q'\right]\\
& \qquad (\epsilon_2<\epsilon_1),
\end{split}
\end{equation}   

\noindent where $q=\epsilon_2/[4\epsilon_1\gamma^2(1-\epsilon_2/\gamma)]$, with $1/(4\gamma^2)<q\leq 1$; and $q'= 4\gamma^2\epsilon_2/\epsilon_1$. Henceforth we ignore the case where $\epsilon_2<\epsilon_1$. \citet{1968PhRv..167.1159J} and \citet{1970RvMP...42..237B} demonstrate that for an isotropic distribution of electrons with a power-law distribution of energy and an index $\delta$ -- that is, $f(\gamma)\sim \gamma^{-\delta}$ -- the resulting spectrum of up-scattered photons is itself a power law with an index $(\delta+1)/2$. Moreover, the up-scattered spectrum is insensitive to the details of the incident photon spectrum. In fact, the energy distribution of the incident photons $\epsilon_1$ may be approximated as a $\delta$ function when $\epsilon_2/\epsilon_1\gg 1$. 

For the particular case of ICS on the Sun we consider, as have previous authors, the case of ultra-relativistic electrons, presumably accelerated by a flare, interacting with soft photospheric photons. Clearly, the photospheric photon field is not isotropic. It is convenient to use the results of \citet{2000ApJ...528..357M}, who derived the differential distribution of up-scattered photons for the more general case of {\sl anisotropic} photons scattering off of an isotropic electron distribution. They find that for an isotropic distribution of mono-energetic electrons described by:

\begin{equation}
f_e(\gamma_e,\Omega_e)=\frac{1}{4\pi\gamma_e^2}\delta(\gamma_e-\gamma),
\end{equation}

\noindent and mono-energetic incident photons with a distribution

\begin{equation}
f_{\gamma}(\epsilon_{\gamma},\Omega_{\gamma})=Q_{\gamma}(\Omega_{\gamma})\frac{1}{\epsilon_\gamma^2}\delta(\epsilon_{\gamma}-\epsilon_1),
\end{equation}

\noindent where $Q_{\gamma}(\Omega_{\gamma})$ is the angular distribution of the photons, the up-scattered photon distribution is given by: 

\begin{equation}\label{eq:int}
\begin{split}
\frac{dR(\gamma)}{d\epsilon_2} = & \frac{\pi r_e^2}{\epsilon_1(\gamma-\epsilon_2)^2}\int_{\Omega_{\gamma}}d \Omega_{\gamma}Q_{\gamma}(\Omega_{\gamma}) \\ 
& \times\left[2-2\frac{\epsilon_2}{\gamma}\left(\frac{1}{\epsilon'_1}+2\right)+\frac{\epsilon_2^2}{\gamma^2}\left(\frac{1}{\epsilon_1^{'2}}+\frac{2}{\epsilon'_1}+3 \right)-\frac{\epsilon_2^3}{\gamma^3}\right],
\end{split}
\end{equation}

\noindent where 

\begin{equation} \label{eq:cond}
\epsilon_2 \leq 2\gamma\epsilon'_1/(1+2\epsilon'_1), \hspace{10 pt} \epsilon'_1=\epsilon_1\gamma(1+\cos\zeta)
\end{equation}

\noindent and $\zeta$ is the angle between the momenta of the electron and the incident photon. We have $\zeta=0$ for a head-on collision from which the maximum energy of the up-scattered photon results:

\begin{equation}
{\epsilon_2}^{max}=\frac{4\epsilon_1\gamma^2}{1+4\epsilon_1\gamma}\approx 4\epsilon_1\gamma^2.
\end{equation}

\begin{figure}
\begin{center}
  \includegraphics[width=0.25\textwidth]{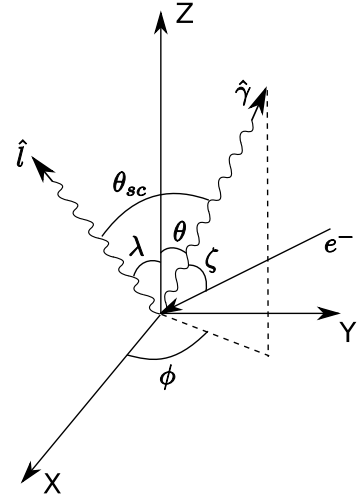}
  \caption{The geometry and angles used to calculate the ICS emission from an anisotropic photon field. The X-Y plane is the solar surface; The Z-axis is normal to the solar surface; $\hat{\gamma}$ is the unit vector along the incident photons, with $\theta$ and $\phi$ as the polar and azimuthal angles; The X-axis is chosen such that the unit vector along the line of sight (LOS) to the observer $\hat{l}$ lies in the X-Z plane; $\lambda$ is the heliocentric angle of the source location; $\theta_{sc}$ is the scattering angle between the incident and up-scattered photons; $\zeta$ is the angle between the electrons and the incident photons.}\label{fig:geometry1}
\end{center}
\end{figure}

\noindent If we assume a fully isotropic photon distribution, i.e., $Q_{\gamma}(\theta,\phi)=1/(4\pi)$, it can be shown \citep{2000ApJ...528..357M} that  Eqn.~ \ref{eq:int}  simplifies to the approximate formula of \citet{1968PhRv..167.1159J} given above by Eqn.~ \ref{eq:jones}. 

To make further progress we adopt the geometry employed by MM10, shown in Fig.~\ref{fig:geometry1}. The Z-axis is normal to the solar surface (the X-Y plane); $\hat{\gamma}$ is the unit vector along the direction of the incident photons, with $\theta$ and $\phi$ as the polar and azimuthal angles; The X-axis is defined to place the unit vector along the line of sight (LOS) to the observer $\hat{l}$ in the X-Z plane, with $\lambda$ the angle between the Z-axis and the LOS (the heliocentric angle of the source from disk center). The scattering angle between the incident photon and the up-scattered photon is given by $\theta_{sc}$. The angle $\zeta$ in the integrand of Eqn.~\ref{eq:int} through Eqn.~\ref{eq:cond} complicates the integration somewhat. Noting that, for ultra-relativistic electrons with $\gamma\gg 1$, incident photons are up-scattered into narrow cone of angular width $\sim 1/\gamma$ in the direction of the energetic electrons, we can approximate the up-scattered photons to be unidirectional which, in turn, allows us to approximate the angle $\zeta$ in Eqn.~ \ref{eq:cond} as $\zeta\simeq \pi-\theta_{sc}$. It is easy to show from the geometry in Fig.~\ref{fig:geometry1} that $\theta_{sc}$ is in fact a function of $\theta$, $\phi$, and $\lambda$: 

\begin{equation}\label{eq:angles}
\cos\theta_{sc}=\cos\theta\cos\lambda+\sin\theta\sin\lambda\cos\phi
\end{equation}

\noindent It is also seen that Eqn.~\ref{eq:cond} requires $\gamma\ge\gamma_{min}=(1/2)\sqrt{\epsilon_2/\epsilon_1}$ in the ultra-relativistic limit. Given the photon distribution function $Q_\gamma(\Omega_\gamma)=Q_{\gamma}(\theta, \phi)$, the integration of Eqn.~\ref{eq:int} becomes straightforward.    

\begin{figure}
\begin{center}
  \includegraphics[width=0.35\textwidth]{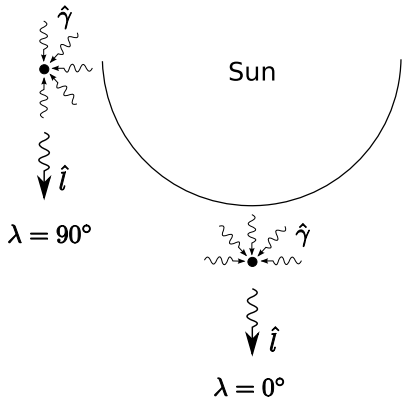}
  \caption{Illustration of the ICS emission for the incident photon distribution in association with the source geometry assumed in MM10.}\label{fig:illust}
\end{center}
\end{figure}

Following MM10, the angular distribution of photospheric photon is taken to fill the half-space above the photosphere (see Fig.~\ref{fig:illust}) and the photon angular distribution is expressed simply as

\begin{equation} \label{eq:pho_dist}
Q_{\gamma}(\theta, \phi)=\frac{1}{2\pi}H(\frac{\pi}{2}-\theta)
\end{equation}   

\noindent where H is the Heaviside function. However, in their subsequent derivation MM10 make an error in the expression for the photon emissivity (their Eqn. 5) that leads to the inclusion of a factor $1+\cos\theta$  rather than $1-\cos\theta_{sc}$. In fact, after substituting the expressions for $Q_{\gamma}(\theta, \phi)$ and $\epsilon'_1$ into Eqn.~\ref{eq:int}, the expression should read

\begin{equation}\label{eq:isoe_urel}
\begin{split}
\frac{dR(\gamma,\lambda)}{d\epsilon_2} = & \left[ \left( 2-\frac{4\epsilon_2}{\gamma}+\frac{3\epsilon_2^2}{\gamma^2}-\frac{\epsilon_2^3}{\gamma^3}\right)\int_0^{2\pi}d\phi\int_0^{\pi/2}\sin\theta d\theta \right. \\
& -\frac{1}{\epsilon_1\gamma}\left( \frac{2\epsilon_2^2}{\gamma^2}-\frac{2\epsilon_2}{\gamma}\right)\int_0^{2\pi}d\phi\int_0^{\pi/2}\frac{d\cos\theta}{1-\cos\theta_{sc}}\\
& \left. -\frac{\epsilon_2^2}{\epsilon_1^2\gamma^4}\int_0^{2\pi}d\phi\int_0^{\pi/2}\frac{d\cos\theta}{(1-\cos\theta_{sc})^2} \right]  \\
& \times\frac{r_e^2}{2\epsilon_1(\gamma-\epsilon_2)^2}
\end{split}
\end{equation}

\noindent where the scattering angle $\theta_{sc}$ is given by Eqn.~ \ref{eq:angles}. The integration limits over $\theta$ also differ from those employed by MM10. Note, too, the kinematic restriction on the scattering angle $\theta_{sc}$ imposed by Eqn.~\ref{eq:cond}:

\begin{equation}
\cos\theta_{sc}\leq 1-\frac{\epsilon_2}{2\epsilon_1\gamma(\gamma-\epsilon_2)}
\end{equation}

\begin{figure*}
\begin{center}
  \includegraphics[width=0.8\textwidth]{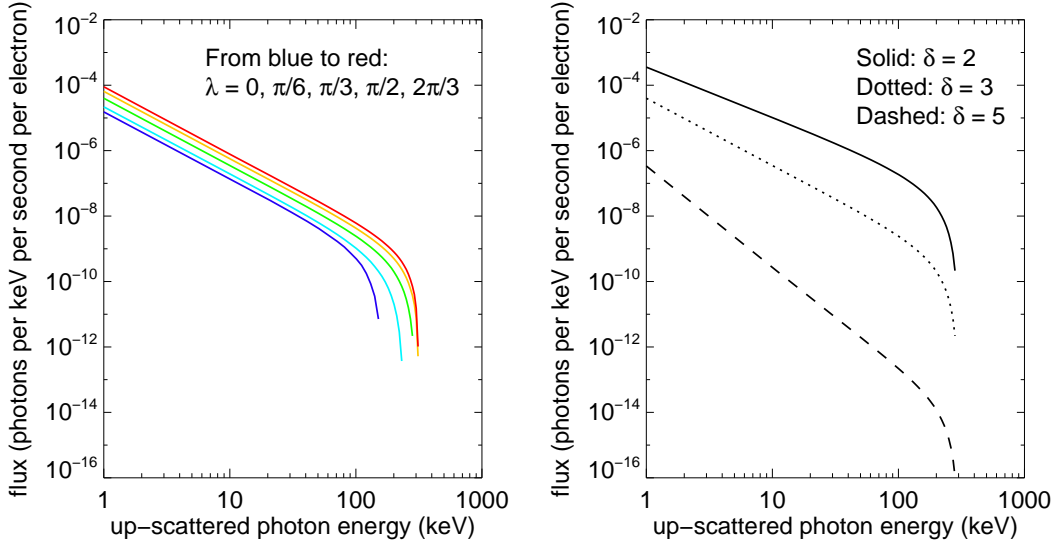}
  \caption{Left: photon flux at the Sun (photons per keV per second per electron) when the source is at disk center ($\lambda=0$; blue), at a longitude of $30^\circ$ ($\lambda=\pi/6$), $60^\circ$ ($\lambda=\pi/3$), $90^\circ$ ($\lambda=\pi/2$), and over-the-limb at $120^\circ$ ($\lambda=2\pi/3$; red). The electron energy spectral index is $\delta=3$. Right: photon flux at the Sun (photons per keV per second per electron) for a source on the limb for the electron energy distribution function with different spectral indexes: solid is for $\delta=2$, dotted for $\delta=3$, and dashed for $\delta=5$. Both panels have an incident photon energy of 2 eV with a number density of $10^{12}$ cm$^{-3}$. }\label{fig:aic_mnm}
\end{center}
\end{figure*}

We have calculated ICS spectra numerically using Eqn.~\ref{eq:isoe_urel}. The photon emissivity spectrum (photons cm$^{-3}$ s$^{-1}$ sr$^{-1}$ keV$^{-1}$) is obtained by integration over the electron and photon energy distributions with suitable normalization. For the purposes of comparison, we use the same parameters as MM10: the incident photons are assumed to be photospheric, with an energy  $\epsilon_1=2$ eV and a number density $n_\gamma=10^{12}$ cm$^{-3}$; the electron kinetic energy is assumed to have a power-law form, $f(\gamma)\sim (\gamma-1)^{-\delta}$. In Fig.~\ref{fig:aic_mnm}a, we show the ICS photon spectra resulting from an electron distribution extending to 100 MeV with a spectral index $\delta=3$, viewed with angles ranging from $\lambda=0$ (disk center) to $\lambda=2\pi/3$ (over-the-limb). The results are normalized such that $n_e(\gamma)=1$ electron cm$^{-3}$ with an energy $>\!0.5$ MeV ($\gamma>2$).  Fig.~\ref{fig:aic_mnm}b shows the ICS spectra from electron distributions with different values of the spectral index $\delta$, for a source on the solar limb ($\lambda=\pi/2$).       

We find that the calculated HXR spectra have a photon spectral index of $\alpha\simeq(\delta+1)/2$, as expected for the ultra-relativistic case \citep{1968PhRv..167.1159J,1970RvMP...42..237B}, similar to those obtained by MM10 (their Fig.~2 and Fig.~4). However, our photon fluxes are more two than orders of magnitude lower than those reported by MM10. They are similar in order of magnitude to the fully-isotropic case calculated from Eqn.~\ref{eq:jones} as might be expected \citep{2000ApJ...528..357M}. We find, moreover, that the difference in the HXR photon spectra calculated for different viewing angles $\lambda$ lie within an order of magnitude of each other, in contrast to the large range of values reported by MM10 which span more than two orders of magnitude.  In practice, the center-to-limb variation of an ICS source would be modified by the contribution of Compton backscatter of ICS HXR photons on photospheric electrons \citep{2006A&A...446.1157K} in the 10$-$100 keV energy range, an effect that we do not include. Note that the high energy cutoff of the photon spectrum depends on viewing angle because the maximum energies from up-scattering are achieved for largest scattering angles. Finally, the high-energy cutoff of the up-scattered photons from different electron power-law energy distributions are independent of the spectral index  $\delta$, while those reported by MM10 vary significantly with $\delta$. 

We conclude that our results for ICS in the limit of ultra-relativistic electron energies are consistent with expectations. We attribute the differences between the calculations reported here and those reported by MM10 to an error made in the expression for the photon emissivity in the latter publication.      

\subsection{Mildly-Relativistic Regime}

We now explore ICS for cases in which the electrons are not necessarily highly relativistic. Solar flares produce copious EUV and SXR photons. These may be up-scattered to HXR or $\gamma$-ray energies by electrons with far lower energies than generally considered by previous treatments of ICS. For example, an $\epsilon_1=1$~keV SXR photon can be up-scattered to $\epsilon_{\rm max}\approx 4\gamma^2\epsilon_1=16-100$~keV for electrons with $\gamma=2-5$. While the photon number density of EUV/SXR photons is much smaller ($\lesssim 10^7 - 10^8$ cm$^{-3}$) than the number density of photospheric photons ($\sim 10^{12}$ cm$^{-3}$) we note that, given power-law, or similar, distributions inferred for electron energy distributions during solar flares, the number of mildly relativistic electrons produced by a solar flare far out-number those at ultra-relativistic energies. The product of the photon number density and the electron number density $n_\gamma n_e$ may therefore not differ substantially between the ultra-relativistic and mildly relativistic cases. 

In considering ICS on mildly relativistic electrons, however, we can no longer exploit the approximations possible for the case of ultra-relativistic electrons and we must instead use the general expression. Consider an electron distribution expressed in separable form as $f_e(\gamma,\Omega)=K_e F_e(\gamma)Q_e(\Omega_e)=K_e F_e(\gamma)/4\pi$ for an isotropic distribution, where $K_e$ is a normalization factor to ensure the integral over the electron energy distribution results in the total number density of fast electrons. The general expression for the ICS emission rate for photons with a direction $\Omega_\gamma$ on an isotropic electron distribution is given by \citet[][see also \citealt{1981Ap&SS..79..321A}]{2000APh....13..107B}

\begin{equation}\label{eq:aic_exact}
\begin{split}
\dfrac{d^2R(\Omega_\gamma)}{d\epsilon_2d\Omega_\gamma} = & \frac{K_e \pi r_0^2 \epsilon_2}{\epsilon_1^2}\int_{\gamma} \left\lbrace \frac{2\epsilon_1}{(\epsilon_1^2+\epsilon_2^2-2\epsilon_1\epsilon_2 \cos\theta_{sc})^{1/2}}\right. \\
+&\left(\frac{1}{R_1}-\frac{1}{R_2} \right) \left[\epsilon_1(1-\cos\theta_{sc})-\frac{2}{\epsilon_2}\right.\\
&-\left.\frac{2}{\epsilon_1 \epsilon_2^2(1-\cos\theta_{sc})}\right] \\
+&\frac{1-\cos\theta_{sc}}{\epsilon_1\epsilon_2^2}\left[\frac{(\gamma-\epsilon_2)\epsilon_2+\gamma\epsilon_1+\epsilon_2\epsilon_1 \cos\theta_{sc}}{R_1^3}\right.\\
+&\left.\left.\frac{(\gamma+\epsilon_1)\epsilon_1+\gamma\epsilon_2-\epsilon_2\epsilon_1 \theta_{sc}}{R_2^3}\right]\right\rbrace \frac{F_e(\gamma)}{\beta\gamma^2}  d\gamma
\end{split}
\end{equation}

\noindent where
\begin{equation}
\begin{split}
R_1=\sqrt{(\gamma-\epsilon_2)^2(1-\cos\theta_{sc})^2+1-\cos^2\theta_{sc}}, \\
R_2=\sqrt{(\gamma+\epsilon_1)^2(1-\cos\theta_{sc})^2+1-\cos^2\theta_{sc}}, 
\end{split}
\end{equation}

\noindent A kinematic limit for $\gamma$ is imposed for any given $\theta_{\rm sc}$, $\epsilon_2$, and $\epsilon_1$. In the Thomson approximation (i.e., $\gamma\epsilon_1 \ll m_e c^2$), the expression is \citep[][Eqn.~ 34]{2000APh....13..107B}:

\begin{equation}
\gamma_{\rm min}=\sqrt{1+\dfrac{(\epsilon_2-\epsilon_1)^2}{2\epsilon_1\epsilon_2(1-\cos\theta_{sc})}}.
\end{equation}

\noindent This kinematic constraint is equivalent to Equation \ref{eq:cond} in the highly relativistic regime.

\begin{figure*}
\begin{center}
  \includegraphics[width=0.85\textwidth]{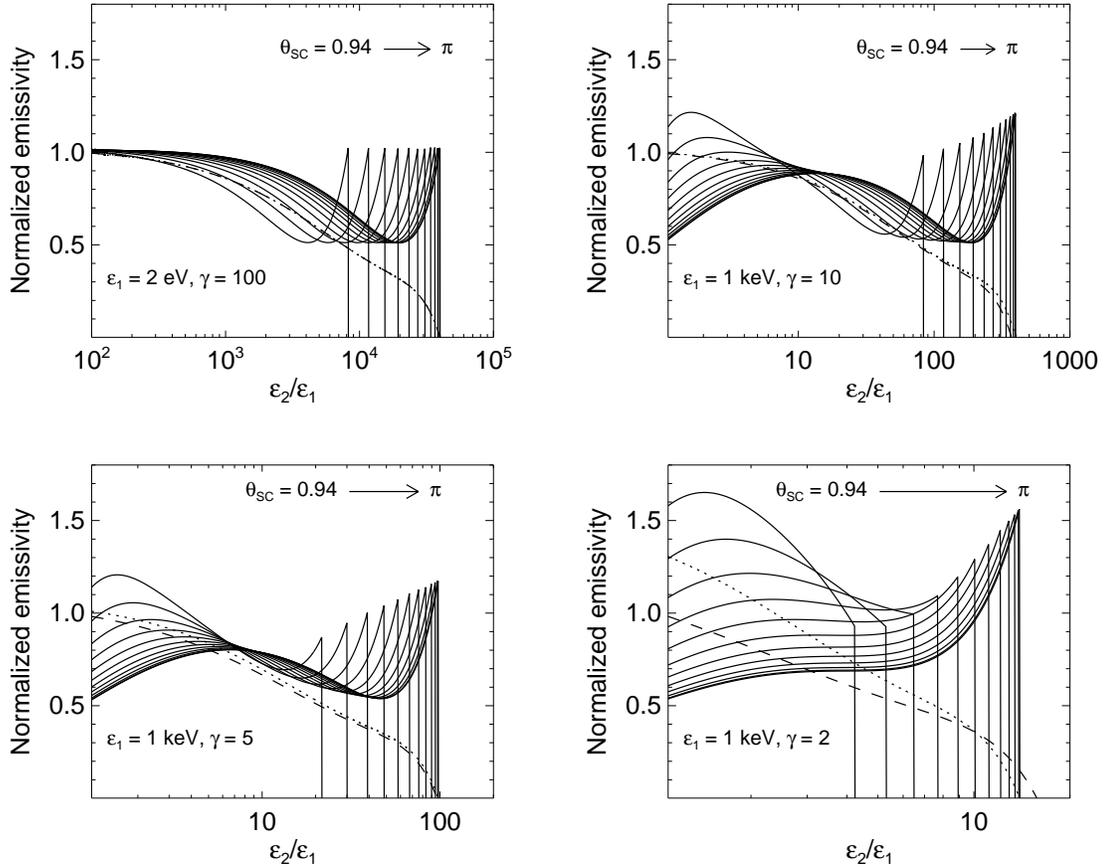}
  \caption{Up-scattered photon spectra from incident photons scattered by an isotropic electron distribution at different scattering angles $\theta_{\rm sc}$ (solid lines) using the exact calculation. The dotted and dashed lines are the results for an isotropic photon distribution by averaging over all the scattering angles, based on the exact formula (Eqn.~ \ref{eq:aic_exact}) and the Jones's formula (Eqn.~ \ref{eq:jones}) respectively. The upper-left panel is for the scattering between the incident photons with $\epsilon_1 = 2$ eV and electrons with $\gamma=100$, i.e., in the ultra-relativistic regime. The other panels are for the mildly relativistic case, with the incident photon energy $\epsilon_1 = 1$ keV and electrons with $\gamma =$ 2, 5 and 10. }\label{fig:spec_angle}
\end{center}
\end{figure*}

Fig.~\ref{fig:spec_angle} shows normalized up-scattered photon spectra for an incident photon scattered by an isotropic electron distribution for  different scattering angles $\theta_{sc}$ calculated using Eqn.~\ref{eq:aic_exact}. We also show the spectrum resulting from an isotropic photon distribution, based on the same equation (dotted line), and from Jones' approximate formula (Eqn.~ \ref{eq:jones}, dashed line). In the ultra-relativistic regime shown in panel (a), the results are similar to those shown in Fig.~2 of \citet{2000ApJ...528..357M}. Note, however, that these authors plot the normalized emissivity as a function of $\epsilon_2/\epsilon_2^{max}$ instead of $\epsilon_2/\epsilon_1$, as we do here.  As $\gamma$ decreases from 100, to 10, to 5 in panels (a)-(c) the spectra at different scattering angles are all peaked at the highest possible energies of the up-scattered photons, with the maximum deviations in emissivity from the isotropic case occurring for $\theta_{sc} \approx \pi$. However, as the electrons enter the mildly relativistic regime ($\gamma$=10, 5, and 2 in panels (b)-(d), respectively) collisions with smaller scattering angles $\theta_{sc}$ contribute significantly to the emissivity at small $\epsilon_2/\epsilon_1$. It is interesting to note that Jones' approximate formula (dashed lines) describes the isotropic case quite accurately for $\gamma \gtrsim 5$, in comparison with the exact calculations (dotted lines). Only at small values of $\gamma$ (panel (d), for which $\gamma=2$) does the spectrum from Jones' approximate formula deviate significantly from the exact calculation for an isotropic photon distribution, both in the spectral shape and the upper cutoff energy.  

\begin{figure*}
\begin{center}
  \includegraphics[width=0.8\textwidth]{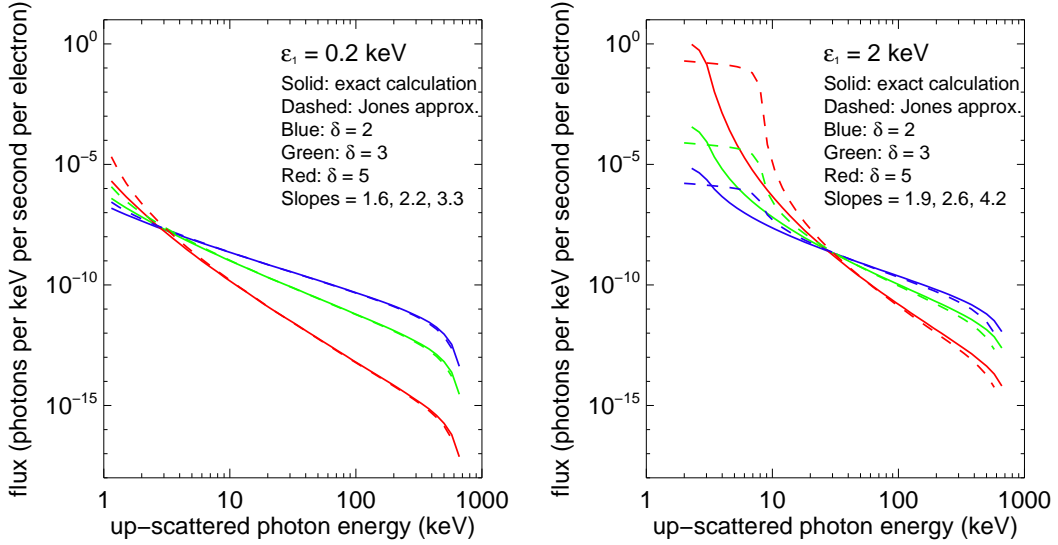}
  \caption{Up-scattered photon spectra at the Sun (photons per keV per second per electron) for the electron energy distribution with different spectral indexes -- $\delta$ = 2 (blue), 3 (green), and 5 (red). The dashed line is the result from Joness approximate formula (Eqn. 1) for the purpose of comparison. The incident photon energy is respectively 0.2 keV and 2 keV in the left and right panel. The incident photon density is assumed to be 10$^7$ cm$^{-3}$. The lower and upper limit to the electron energy spectrum is $\gamma_1=1.02$ ($\approx$ 10 keV, both panels), $\gamma_2=30$ ($\approx 14.8$ MeV, left panel), and 10 ($\approx$ 4.6 MeV, right panel). The slopes are obtained by fitting a power law between 20 keV and 80 keV.  }\label{fig:isoe_mrel}
\end{center}
\end{figure*}

Adopting the same geometry as shown in Fig.~\ref{fig:geometry1}, the resulting ICS flux can be obtained by integrating $d^2R(\Omega_\gamma,\epsilon_1)/d\epsilon_2d\Omega_\gamma$ over the solid angle $\Omega_{\gamma}$ of any incident photon distribution $Q_{\gamma}(\theta, \phi)$, i.e.:

\begin{equation}\label{eq:int_omega}
\dfrac{dR(\lambda)}{d\epsilon_2}=\int_0^{2\pi} d\phi \int_0^{\pi} \sin\theta d\theta Q_\gamma(\theta,\phi) \frac{d^2R(\Omega_\gamma)}{d\epsilon_2d\Omega_\gamma}
\end{equation}
\noindent Fig.~\ref{fig:isoe_mrel} shows up-scattered photon spectra resulting from an isotropic distribution (i.e., $Q_{\gamma}=1/(4\pi)$) of mono-energetic photons interacting with an isotropic distribution of electrons. The incoming photon energy is assumed to be $\epsilon_1=0.2$ keV (left panel) and $\epsilon_1=2$ keV (right panel) with a photon density of $10^7$ cm$^{-3}$. The electron kinetic energy has a power-law form $\sim (\gamma-1)^{-\delta}$ with $\delta=3$. The lower and upper limits are assumed to be $\gamma_1=1.02$ (10 keV), $\gamma_2=30$ ($\approx 14.8$ MeV, left panel), and $\gamma_2=10$ ($\approx 4.6$ MeV, right panel). The results are normalized to one electron above $\sim 0.5$ MeV as was done previously (\S2.1). 

The main difference between the ultra-relativistic case and the mildly relativistic case is that spectra in the latter regime are significantly steeper than the classic relation $\alpha\sim(\delta+1)/2$. Jones' approximate formula (Eqn.~ \ref{eq:jones}) describes the up-scattered photon spectra quite well (the thin-dashed curves) at most photon energies $\epsilon_2$, the largest deviations occurring at the lowest ratios of $\epsilon_2/\epsilon_1$.

Our results for ICS of an anisotropic photon field on an isotropic distribution of mildly relativistic electrons, not shown here, are qualitatively similar to those reported in \S2.1 for the ultra-relativistic case (Fig.~\ref{fig:aic_mnm}),. The main difference in this case is that while the emissivity increases with increasing values of $\lambda$, the variation from $\lambda=0$ to $\lambda=\pi/2$ is somewhat less than is seen for the ultra-relativistic case for $\epsilon_2/\epsilon_1\gtrsim 10-20$. For smaller values of $\epsilon_2/\epsilon_1$ the spectra become relatively insensitive to $\lambda$. This is not unexpected because at lower electron energies the emission pattern becomes quite broad and the directionality of the source therefore decreases. 

\section{ICS of Isotropic Photons on Anisotropic Electron Distributions}

Anisotropic electron distributions may arise during solar flares as a result of acceleration, injection, and/or transport effects. Two idealized cases are commonly considered: electrons in a beam or electrons in a plane. We refer to the latter type as a ``pancake" distribution. In the presence of a magnetic field, beamed electrons are typically those with pitch angles $\alpha_p$ such that $u=\cos\alpha_p$ is confined to be near 1 or -1 (i.e., they stream along the magnetic field) and pancake distributions are those where $u\sim 0$ (i.e., electrons are  largely perpendicular to the magnetic field).   We consider both of these distributions in this section, and calculate the ICS spectra resulting from their interaction with an isotropic photon field. We use an isotropic photon field for computational ease but note that the effect of an anisotropic photon field manifests itself largely as a geometrical effect resulting in a correction factor of order $4\pi/\Omega_\gamma$ \citep{2000ApJ...528..357M}. For ICS of photospheric photons scattering on ultra-relativistic regime, the effect is therefore a factor of about 2. 

\begin{figure}
\begin{center}
  \includegraphics[width=0.25\textwidth]{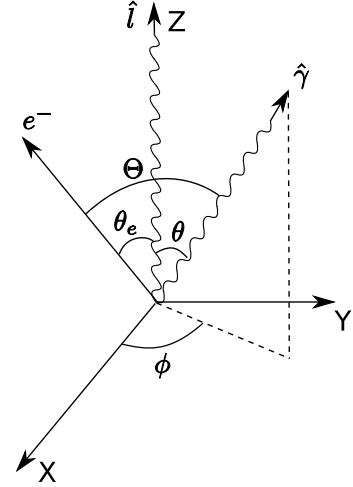}
  \caption{The geometry and angles used to calculate the ICS emission between a uni-directional photon beam and a uni-directional electron beam, each normalized to unit density. The Z-axis is along the direction of the up-scattered photons (i.e., the line of sight $\hat{l}$); $\hat{\gamma}$ is the unit vector along the incident photon beam, with $\theta$ and $\phi$ as the polar and azimuthal angles; The direction of the electron is placed in the X-Z plane, with an angle $\theta_e$ away from the LOS; $\Theta$ is the angle between the electron and the incident photon. }\label{fig:geometry2}
\end{center}
\end{figure}  

\citet{2000APh....13..107B} derived the ICS emissivity from unidirectional, mono-energetic photons interacting with a unidirectional electron distribution with an arbitrary energy distribution. We adopt these results here but transform to a more convenient geometry, as shown in Fig.~\ref{fig:geometry2}. In our geometry, the line of sight is along the Z-axis which is taken to be the direction of the up-scattered photon ($\hat{l}$ is the unit vector). The incident photons have a direction of $\Omega_\gamma(\theta,\phi)$, where $\theta$ and $\phi$ are the polar and azimuthal angles respectively. The direction of the electron momenta is taken to be in the X-Z plane, with an angle $\theta_e$ away from the LOS. $\Theta$ is the angle between the electron and the incident photon. The derived emission rate is then (adapted from Eqn. (26) of \citealt{2000APh....13..107B}):

\begin{equation}
\begin{split}
\dfrac{d^3R(\Omega_\gamma,\theta_e)}{d\epsilon_2d\Omega_\gamma d\Omega_e}=& 2\pi r_0^2  \frac{\epsilon_2}{\epsilon_1}\frac{(\cos\theta-1)^2}{\epsilon_2-\epsilon_1}\left\lbrace \frac{1}{(\cos\theta-1)^2} \right.\\
&+\left.\left[\frac{1}{\cos\theta-1}+\frac{(\epsilon_2-\epsilon_1)^2}{\epsilon_1\epsilon_2(\cos\theta_e-\cos\Theta)^2}\frac{1}{\gamma_0^2} \right]^2\right\rbrace \\
&\times\left(\gamma_0-\frac{1}{\gamma_0}\right)K_e F_e(\gamma_0)Q_e(\Omega_e)Q_{\gamma}(\Omega_\gamma), 
\end{split}
\end{equation} 

\noindent where the Lorentz factor $\gamma_0=1/\sqrt{1-\beta_0^2}$ of the electron is determined by kinematics

\begin{equation}
\beta_0=\frac{\epsilon_2-\epsilon_1}{\epsilon_2\cos\theta_e-\epsilon_1\cos\Theta},
\end{equation}

\noindent $K_e F_e(\gamma_0)Q_e(\Omega_e)$ is the differential electron number density at $\gamma_0$, and $\Theta$ is related to $\theta$, $\phi$, and $\theta_e$ by geometry

\begin{equation}
\cos\Theta=\cos\theta\cos\theta_e+\sin\theta\sin\theta_e\cos\phi.
\end{equation}

\noindent For an isotropic incident photon field $Q_{\gamma}(\theta, \phi)=1/(4\pi)$, integration over the solid angle $\Omega_{\gamma}$ yields the ICS rate $d^2R(\theta_e)/d\epsilon_2 d\Omega_e$ for a given direction of the electron $\Omega_e$.    

\begin{figure}
\begin{center}
  \includegraphics[width=0.4\textwidth]{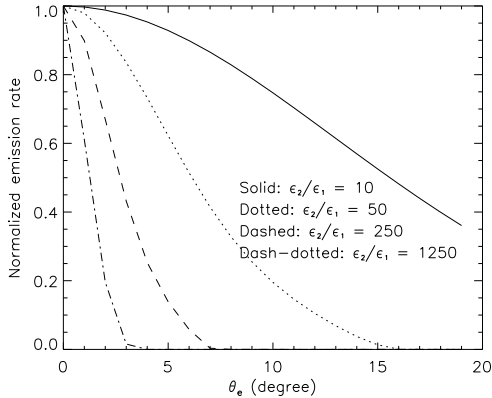}
  \caption{Normalized ICS emission rate at different angles of the electron relative to the LOS ($\theta_e$). The solid, dotted, dashed, and dash-dotted curves are for $\epsilon_2/\epsilon_1=$ 10, 50, 250, and 1250 respectively.}\label{fig:beam_angle}
\end{center}
\end{figure}

Fig.~\ref{fig:beam_angle} shows the normalized ICS emission rates as a function of the source electron angle $\theta_e$ at different ratios of  $\epsilon_2/\epsilon_1$. As expected, the emission is most favorable when the electron direction is along the LOS (i.e., $\theta_e=0$), and drops rapidly with increasing $\theta_e$. In addition, the emission cone opens up gradually from the ultra-relativistic case (large $\epsilon_2/\epsilon_1$) to the mildly relativistic case (smaller $\epsilon_2/\epsilon_1$). The width of the emission cone is $\sim 1/\gamma_{\rm min}$, where $\gamma_{\rm min}\approx \frac{1}{2}\sqrt{\epsilon_2/\epsilon_1}$ is the minimum electron Lorentz factor required to up-scatter $\epsilon_1$ to $\epsilon_2$.  

\begin{figure*}
\begin{center}
  \includegraphics[width=0.8\textwidth]{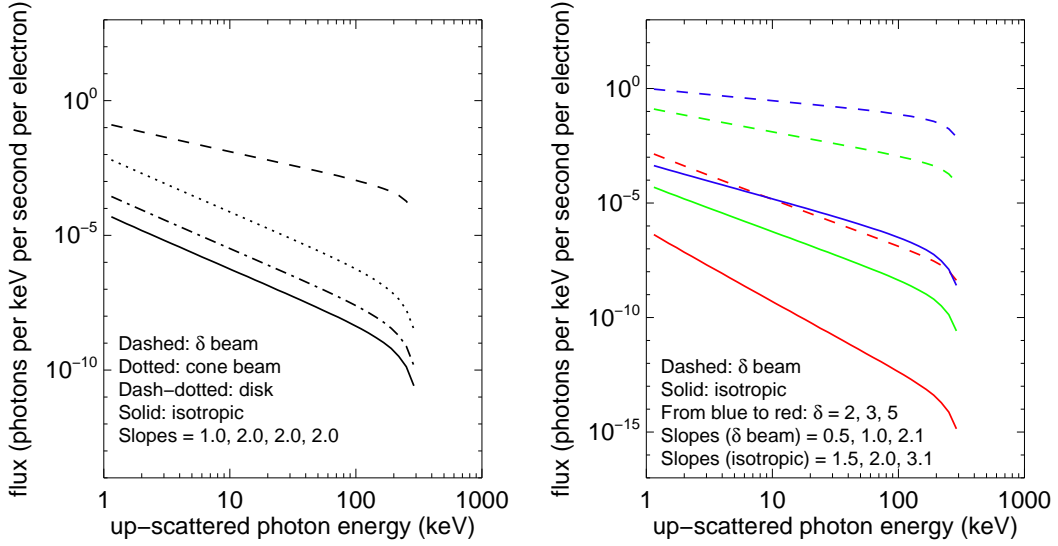}
  \caption{Up-scattered photon spectra at the Sun (photons per keV per second per electron) from anisotropic electron distributions along the line of sight in the ultra-relativistic regime. Left: comparison between the spectra from a mono-directional electron beam (dashed line), a cone-like distribution of electrons with a half width $\Delta \theta_b = 10^{\circ}$ (dotted line), a pancake electron distribution with a half-angle width of $\Delta \theta_p=10^{\circ}$ (dash-dot line), and an isotropic electron distribution (solid line). The electron energy spectral index is $\delta=3$. Right: the ICS spectra resulting from the $\delta$-function beam (dashed) and isotropic (solid) distributions with different electron spectral indexes $\delta$. The blue, green, and red curves are for $\delta=$2, 3, and 5 respectively. All of the curves assume an incident photon energy of 2 eV, and the number density is assumed to be $10^{12}$ cm$^{-3}$.}\label{fig:beam_urel}
\end{center}
\end{figure*}

The ICS emissivity can be obtained by integrating over the solid angle $\Omega_e$ for any given electron distribution $Q_e(\Omega_e)$

\begin{equation} \label{eq:int_omega_e}
\frac{dR}{d\epsilon_2}=\int d\Omega_e \frac{d^2R(\theta_e)}{d\epsilon_2 d\Omega_e} K_e F_e(\gamma_0) Q_e(\Omega_e) . 
\end{equation}

\begin{figure*}
\begin{center}
  \includegraphics[width=0.8\textwidth]{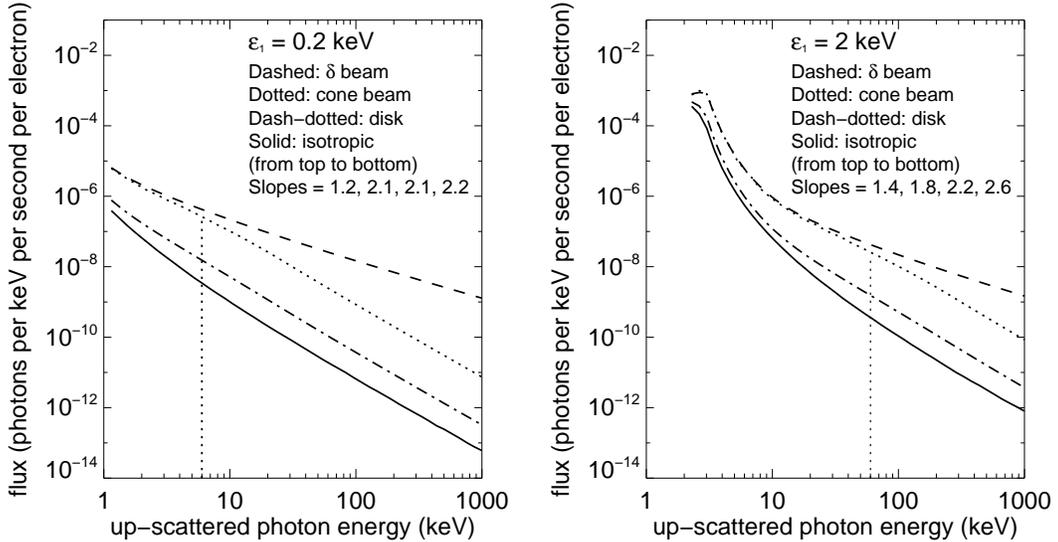}
  \caption{Up-scattered photon spectra at the Sun (photons per keV per second per electron) from anisotropic electron distributions along the line of sight in the mildly relativistic regime. Left: comparison between the spectra from different types of electron angular distributions. The dashed, dotted, dashed-dotted, and solid curves are respectively for a mono-directional beam, a cone-beam (with a half width $\Delta \theta_b = 10^{\circ}$), a pancake electron distribution (with a half-angle thickness of $\Delta \theta_p=10^{\circ}$), and an isotropic electron field. The incident photons have an energy of 0.2 keV with a number density of $10^{7}$ cm$^{-3}$, and the electron energy spectral index is $\delta=3$. Right: all parameters are the same except the incident photon energy is 2 keV. The spectral slopes are obtained by fitting the spectra between 20$-$80~keV. The vertical dotted lines indicate the location of $(\epsilon_2/\epsilon_1)_0=30$ that corresponds to the transition of the broken power-law in the spectra of the cone-beam distribution.}\label{fig:beam_mrel}
\end{center}
\end{figure*}

\noindent The most favorable and extreme case is for a mono-directional electron beam along the line of sight.  In the ultra-relativistic regime, the ICS emission rates (Fig.~\ref{fig:beam_urel}, dashed curves) are 4$-$5 orders of magnitude higher than the isotropic case from 10$-$100 keV (Fig.~\ref{fig:beam_urel}, solid curves). The spectra are significantly flatter, with a photon spectral index, $\alpha \sim (\delta-1)/2$, similar to the analytical results in \citet{2000APh....13..107B} (see his Eqn.~28; note that his result is expressed in terms of an energy emissivity rather than a photon emissivity). In the mildly relativistic regime (Fig.~\ref{fig:beam_mrel}, dashed curves), the ICS photon spectrum is likewise significantly flatter than the isotropic case, although it is somewhat steeper than the ultra-relativistic case. Here, too, the ICS emission is enhanced by  over the isotropic case: by 2.5$-$3.5 orders of magnitude for EUV photons and by 1$-$2.5 orders of magnitude for SXR photons. 

A more physically realistic beamed electron distribution is not mono-directional, but has a finite angular width. We therefore consider electrons confined to a cone with an angular half-width of $\Delta\theta_b$ with its axis along the line of sight:

\begin{equation}
Q_e(\Omega_e)=\frac{H(\Delta \theta_b-\theta_e)}{2\pi(1-\cos\Delta \theta_b)},
\end{equation}

\noindent Results are again calculated in both the ultra-relativistic (Fig.~\ref{fig:beam_urel}, dotted curve) and mildly relativistic regimes (Fig.~\ref{fig:beam_mrel}, dotted curves). For the ultra-relativistic case, the resulting spectra are steeper than that resulting from the $\delta$-function beam and, in fact, have the same shape as the isotropic case, but differ in magnitude by a constant factor. The reason is that since the width of the electron emission cone goes as $1/\gamma$, electrons with smaller $\gamma$ have broader emission cones than those with larger $\gamma$. Lower energy electrons therefore contribute to an increasing degree as $\theta$ deviates from the LOS, resulting in a steeper spectrum. In fact, there should be no significant difference from the isotropic case except for a geometric filling factor such that the emissivity is enhanced by a factor $\sim\!4/\Delta\theta_b^2$ for small values of $\Delta\theta_b$. 

The ICS spectra in the mildly relativistic regime present more complex features (Fig.~\ref{fig:beam_mrel}): they are similar to the spectra of the mono-directional beam case at small values of $\epsilon_2/\epsilon_1$, and transition to a form similar to that obtained for the ultra-relativistic case at large values of $\epsilon_2/\epsilon_1$. The spectra resemble a broken power-law that is harder at small $\epsilon_2/\epsilon_1$ and softer at large $\epsilon_2/\epsilon_1$. This can also be understood by the variation of the emission cone with $\epsilon_2/\epsilon_1$ -- when $\epsilon_2/\epsilon_1$ is small, the emission cone is much wider than the beam width $\Delta \theta_b$. That means all the electrons in the cone-beam contribute near equally to the ICS flux, so the spectra are similar to those of the $\delta$-function beam case. As $\epsilon_2/\epsilon_1$ gets larger, the width of the emission cone decreases and the emission asymptotically approaches the ultra-relativistic regime. The transition happens when the half-width of the emission cone $1/(2\gamma_{\rm min})\approx \sqrt{\epsilon_1/\epsilon_2}$ reaches the half-width of the cone-beam $\Delta \theta_b$, which sets a critical value of $(\epsilon_2/\epsilon_1)_0\approx 1/\Delta \theta_b^2$ for the transition. In Fig.~\ref{fig:beam_mrel}, where $\Delta \theta_b = 10^{\circ}$, the transition occurs where $(\epsilon_2/\epsilon_1)_0\approx 30$, marked by the vertical dotted lines in Fig.~\ref{fig:beam_mrel}.      

\begin{figure}
\begin{center}
  \includegraphics[width=0.3\textwidth]{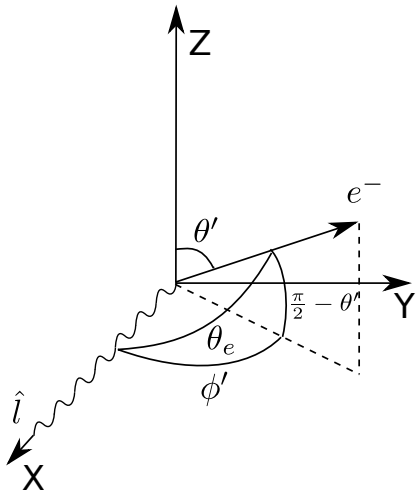}
  \caption{The geometry and angles used to calculate the ICS emission between a disk-like electron angular distribution and an isotropic photon field. The disk is in the X-Y plane. The X-axis is along the line of sight $\hat{l}$. $\theta'$ and $\phi'$ are the polar and azimuthal angles of the electron direction. $\theta_e$ is the angle between the electron and LOS. The electrons are confined in polar angles from $\pi/2-\Delta \theta_p$ to $\pi/2+\Delta \theta_p$, where $\Delta \theta_p$ is the half-angular-thickness of the disk.}\label{fig:geo_disk}
\end{center}
\end{figure}

We now consider a pancake electron anisotropy. In the case of a magnetized plasma, such a distribution arises when the electron pitch angles are largely perpendicular to the magnetic field ($u\sim 0$). Fig.~\ref{fig:geo_disk} describes the geometry for calculating the ICS emission with a pancake electron angular distribution: the electrons are taken to be confined near the X-Y plane, with the LOS along the X-axis. The direction of the electron $\Omega_e$ is described by the polar and azimuthal angles $\theta'$ and $\phi'$. The electrons are confined in a polar angles from $\pi/2-\Delta \theta_p$ to $\pi/2+\Delta \theta_p$, where $\Delta \theta_p$ is the half-thickness of the disk. The electron angular distribution $Q_e(\Omega_e)$ is assumed to be

\begin{equation}
Q_e(\Omega_e)=\frac{H\left(\Delta \theta_p-\left|\theta'-\pi/2\right|\right)}{4\pi\sin\Delta \theta_p},
\end{equation}

\noindent where we again have $\int Q_e(\Omega_e) d\Omega_e=1$. The angle between the electron and the LOS $\theta_e$ is related to the polar and azimuthal angles of the electron direction ($\theta'$, $\phi'$) by

\begin{equation}
\cos \theta_e=\cos\phi' \sin\theta'.
\end{equation}
\noindent Hence, the ICS emissivity can be obtained by integrating $d^2R(\theta_e)/d\epsilon_2 d\Omega_e$ over the solid angle $\Omega_e$ in Eqn.~\ref{eq:int_omega_e}.

The results for both the ultra- and mildly relativistic cases are shown in Fig.~\ref{fig:beam_urel} and \ref{fig:beam_mrel} as dash-dotted curves, using the same parameters as those in the mono-directional and cone-beam cases, with the half-thickness of the disk $\Delta \theta_p=10^{\circ}$. The spectra are similar to those of the cone-beam, except that the magnitude is now enhanced by a factor of $\sim 1/\Delta \theta_p$ for small $\Delta \theta_p$ - e.g., about a factor of 6 when $\Delta\theta_p=10^\circ$ - compared to the fully-isotropic case.  There is less obvious complexity to the photon spectrum resulting from a pancake electron distribution. The reasons is that in the plane of the pancake, there is no cutoff effect with increasing $\epsilon_2/\epsilon_1$; it is only present in the polar angle, perpendicular to the pancake. Since, in the mildly relativistic case, the photon spectrum steepens for smaller values of $\epsilon_2/\epsilon_1$, the effect manifests itself subtly in Fig.~\ref{fig:beam_mrel} as a convergence between the dotted curves (cone-beam anisotropy) and the solid line curves (isotropic case) for $\epsilon_2/\epsilon_1<(\epsilon_2/\epsilon_1)_0$; i.e, the photon spectrum does not steepen as much as it otherwise would.

In summary, ICS on anisotropic electron distributions can be far more effective than it is on an isotropic electron distribution. The reason is that ICS is highly directional - those electrons that are far from the line of sight do not contribute to ICS. As a result, all other things being equal, an anisotropic electron distribution results in significantly enhanced ICS emission for favorable viewing geometries; i.e., when the electron momenta are near the line of sight. 

\section{Thin-Target Non-thermal Bremsstrahlung Emission}

In this section we briefly review results for thin-target non-thermal bremsstrahlung emission and compare this mechanism with ICS. As noted in \S1, coronal HXR sources have been interpreted in terms of nonthermal bremsstrahlung emission in general. Since the plasma density in the corona is low the ``thin-target'' regime is usually assumed. There are exceptions, of course: collisionally thick coronal HXR sources may occur as a result of a high ambient plasma density \citep{2004ApJ...603L.117V} or trapping \citep[e.g.,][]{2007ApJ...669L..49K}. We only consider the thin-target regime here. 

The total bremsstrahlung cross-section is composed of two parts: that due to collisions between incident electrons and target ions, and that due to collisions between incident electrons electrons and target electrons. The doubly differential electron-ion bremsstrahlung (EIB) cross-section is given by \citet{1959RvMP...31..920K} (equation 2BN) whereas the electron-electron bremsstrahlung (EEB) cross-section is given by \citet{1975ZNatA..30.1099H}. The contribution of EEB to the total bremsstrahlung emission is negligible at non-relativistic electron energies. For this reason it is typically ignored when considering HXR photon energies less than a few $\times 100$ keV and the emission is assumed to be the result of EIB alone. However, as emphasized by \citet{2007ApJ...670..857K}, the neglect of EEB is not necessarily justified when considering photon energies $\gtrsim300$ keV. For an isotropic distribution of non-relativistic electrons with a power-law density distribution and a spectral index $\delta$, the HXR photon spectral index due to thin-target EIB is $\alpha\sim \delta+1/2$.   For incident electrons approaching relativistic energies, the thin-target EIB spectrum flattens and $\alpha \sim \delta$. The EEB contribution has a photon spectral index $\alpha\sim \delta-1/2$ at lower energies and flattens somewhat (by 0.2$-$0.3) at higher energies. Hence, the EEB contribution can further flatten the total photon spectrum as photon energies exceed a few hundred keV. Previous work \cite[e.g., ][]{1975SoPh...45..453H} has shown, however, that for isotropic distributions of incident electrons on a pure hydrogen plasma, the contribution of EEB is significantly less than EIB. For example, EEB contributes roughly 27\%, 18\% and 12.5\% of the total photon emissivity at 400 keV for emission from a power-law distribution of energetic electrons with $\delta=3, 4,$ and 5, respectively. The EEB contribution results in an corresponding hardening of the photon spectral index in the amount of 0.2, 0.15, and 0.1, respectively, for photon energies $>200$ keV. 

\begin{figure*}
\begin{center}
 \includegraphics[width=0.9\textwidth]{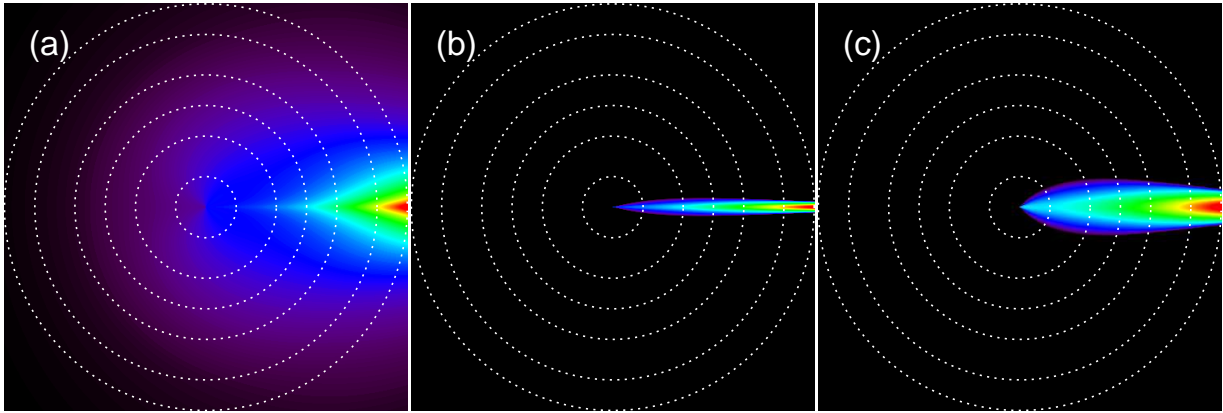}
  \caption{Polar diagram of the normalized emissivity from a mono-directional electron as a function of the logarithm of the photon energy $\epsilon_\gamma$ and the angle of the line of sight relative to the beam direction. The electron beam propagates in the $+x$ direction. The electron distribution function is a power law between 10 keV and 100 MeV and has an index $\delta=3$. The emissivity has been multiplied by $\epsilon_\gamma^\delta$ in all cases. The concentric circles indicate contours of constant photon energy: 20, 50, 100, 200, 500, and 1000 keV. a) Thin-target bremsstrahlung emissivity. The EIB and EEB contributions have been summed;  b) the same for ICS of mono-energetic EUV photons (0.2 keV) up-scattered by the electron beam; c) same as panel (b), but for SXR photons (2 keV).}\label{fig:aic_abr_angle}
\end{center}
\end{figure*}

The bremsstrahlung emissivity, like ICS, is anisotropic. Fig. \ref{fig:aic_abr_angle} shows a polar diagram of the normalized thin-target bremsstrahlung emissivity (EIB+EEB) as a function of $\theta_e$  and the logarithm of the photon energy resulting from a mono-directional ($\delta$-function) angular distribution and a power-law energy distribution of electrons ($\delta=3$) extending from 10 keV to 100 MeV. It is compared with ICS from the same distribution of electrons scattering EUV photons ($\epsilon_1=0.2$ keV), and SXR photons ($\epsilon_1=2$ keV). The case of photospheric photons (2 eV) scattering from the ultra-relativistic electrons in the distribution is not shown because the emissivity is itself mono-directional. Similar to the ICS emissivity, the bremsstrahlung emissivity peaks in the direction of the incident electrons at large photon energy $\epsilon_2$. Its emission pattern is more complex at non-relativistic energies, showing a maximum at $\theta_e\sim 30$-$40^\circ$ rather than along the line of sight \citep[cf., ][]{2004ApJ...613.1233M}. In any case, the bremsstrahlung emission pattern is broader than that of ICS as a result of the fact that the energy of a given bremsstrahlung photon is comparable to that of the incident electron ($\gamma\sim \epsilon$) whereas the electrons responsible for ICS are much more energetic than the up-scattered photons considered here ($\gamma\gg \epsilon$) and their beaming is consequently more pronounced. 

For completeness, we have also considered the effect of a low energy cutoff $E_c$  in the electron energy distribution on the spectral index of the photon spectrum for EIB, EEB, and their sum. Our results are consistent with those of \citet{2008A&A...486.1023B} for the case of an isotropic electron distribution and thin target EIB emission. That is, for photon energies well below the cutoff ($\epsilon\ll E_c$) the spectral index is very hard ($\alpha\sim 1.5$).  The inclusion of EEB may change the spectral index of the photon spectrum by $\lesssim 0.1$. 

\begin{figure}
\begin{center}
 \includegraphics[width=0.4\textwidth]{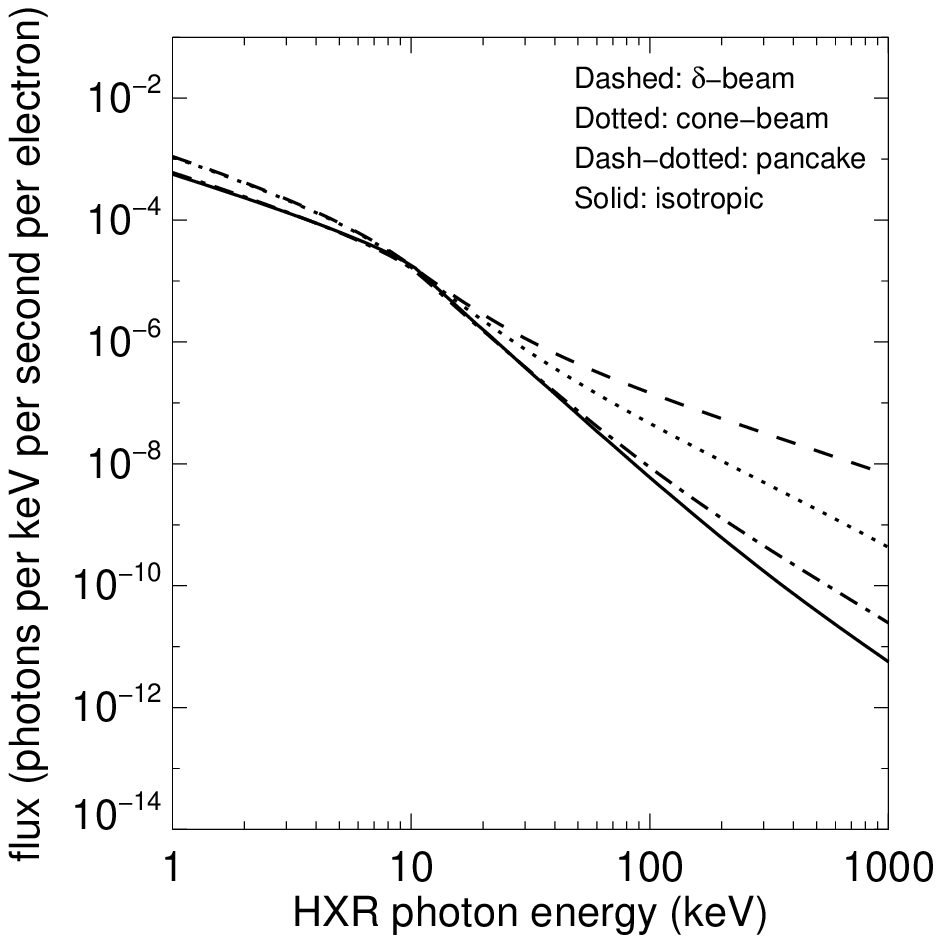}
\caption{Bremsstrahlung photon spectra at the Sun (photons per keV per second per electron) from anisotropic electron distributions along the line of sight. Different kinds of anisotropies are plotted - a mono-directional electron beam (dashed line), a cone-beam distribution with a half-angle width of $\Delta \theta_b=10^{\circ}$ (dotted line), a pancake electron distribution with a half-angle width of $\Delta \theta_p=10^{\circ}$ (dash-dot line), and an isotropic electron distribution (solid line). The electron energy distribution is assumed to have a single power-law form with a spectral index $\delta=3$, extending from 10 keV to 100 MeV (same as that in Fig. \ref{fig:beam_mrel}). The results are also normalized to one source electron above 0.5 MeV, and the ion number density $n_i$ is assumed to be $10^8$ cm$^{-3}$. 
Note the break in the spectra at $\sim$ 10 keV is from the lower energy cutoff of the electron distribution at 10 keV.}\label{fig:abr_emis}
\end{center}
\end{figure}

We have also calculated the bremsstrahlung photon spectra, including the contributions from both EIB and EEB, resulting from the same electron anisotropies considered in \S3, namely, a mono-directionall electron beam, a cone-beam distribution with a half-angle width of $\Delta \theta_b=10^\circ$, and a pancake electron distribution with a half-angle width of $\Delta \theta_p=10^\circ$. Fig. \ref{fig:abr_emis} shows the corresponding results from a single power-law electron energy distributions with a spectral index $\delta=3.0$ using the same parameters used to compute the examples of ICS emission in the mildly-relativistic regime shown in Fig. \ref{fig:beam_mrel}.  We note that the resulting spectra are qualitatively similar to those resulting from ICS on these distributions. In particular, the extreme case of a mono-directional electron beam directed along the line of sight results in a significant enhancement to the thin-target emissivity and a substantially flatter spectrum than the isotropic case, for which $\alpha=3.5$ for non-relativistic photon energies (20$-$80 keV) and $\alpha=2.9$ for $\gamma$-ray photon energies (200$-$800 keV).  The cone-beam and pancake anisotropy result in more modest enhancements and their spectra are intermediate to the isotropic and mono-directional beam (for the cone beam, $\alpha=$2.5 and 2.0 for the 20$-$80 keV and 200$-$800 keV ranges respectively; for the pancake anisotropy, $\alpha=$3.2 and 2.5, respectively). We note that for the mono-directional beam, EIB and EEB asymptotically approach equality as the photon energy increases \citep[cf., ][]{1986ApJ...301..962D}, but the EEB contribution becomes less prominent in the beam-cone and pancake distributions. Note, too, that the degree of enhancement of each of the anisotropic cases relative to the isotropic case is less dramatic than for ICS. There is essentially no enhancement at 10 keV photon energies owing to the fact that the electron distribution cuts off at 10 keV, but this changes as the photon energy increases: to an enhancement of perhaps 1.5 orders of magnitude for the mono-directional beam at 100 keV; $\lesssim 1$ order of magnitude for the cone beam; and a factor of $\sim 2$ for the pancake distribution. This can be understood as a consequence of the more modest degree of directivity of bremsstrahlung emission compared with ICS.  We conclude from this exercise that the effect of electron anisotropies on mildly relativistic and ultra-relativistic ICS emission is significantly larger than is the case for thin-target bremsstrahlung emission, all other things being equal. 

\begin{deluxetable*}{rrrrrrrrrrrrrrr}
\tablecolumns{15}
\tablewidth{0pc}
\tablecaption{Comparison of Fitted HXR/$\gamma$-ray Photon Spectral Indices}
\tablehead{
\colhead{}    &  \multicolumn{4}{c}{UR ICS} & \colhead{} & \multicolumn{4}{c}{MR ICS} &
\colhead{} & \multicolumn{4}{c}{Bremsstrahlung} \\
\cline{2-5} \cline{7-10} \cline{12-15}\\
\colhead{keV} & \colhead{ISO}   & \colhead{PAN}    & \colhead{C-B} & \colhead{$\delta$-B} &
\colhead{} & \colhead{ISO}   & \colhead{PAN}    & \colhead{C-B} & \colhead{$\delta$-B} &
\colhead{} & \colhead{ISO}   & \colhead{PAN}    & \colhead{C-B} & \colhead{$\delta$-B} }
\startdata
20$-$80 & 2.0 & 2.0 & 2.0 & 1.0 & & 2.6 & 2.2 & 1.8 & 1.4 & & 3.5 & 3.2 & 2.5 & 1.9 \\
200$-$800 & 2.0 & 2.0 & 2.0 & 1.0 & & 2.1 & 2.1 & 2.1 & 1.1 & & 2.9 & 2.5 & 2.0 & 1.3 \\
\enddata
\tablecomments{ISO: isotropic, PAN: pancake, C-B: cone-beam, $\delta$-B: mono-directional beam. The results are calculated based on a single power-law electron energy distributions with a spectral index $\delta=$3. For the mildly-relativistic ICS case, an incident photon energy of 2 keV is used.}
\end{deluxetable*}

We now turn our attention to the relative roles of ICS and thin-target bremsstrahlung in the production of HXR and continuum $\gamma$-ray emission. As was discussed by MM10, a comparison between the relative roles of ICS and bremsstrahlung is somewhat problematic because the HXR photons resulting from the two mechanisms are due to electrons from very different parts of the electron energy distribution. The high energy electrons responsible for ICS make essentially no contribution to the HXR bremsstrahlung emission. Similarly, the much lower energy electrons responsible for HXR bremsstrahlung emission contribute no ICS emission. Nevertheless, \citet{1971SoPh...18..284K} compared the two mechanisms for a single isotropic, power-law electron distribution for ICS in the ultra-relativistic limit.  This result was recast by \citet{2008A&ARv..16..155K} as the ratio $R$ of the ICS to EIB emissivities, given approximately as:

\begin{equation}
R=\frac{3}{2\alpha'}\frac{n_\gamma}{n_p} (2\delta-1)Q(\delta)\Bigl(\frac{\epsilon_2} {4\epsilon_1}\Bigr)^{(1-\delta)/2}\Bigl(\frac{\epsilon_2} {m_e c^2}\Bigr)^{\delta-1/2}
\end{equation}
where $n_\gamma$ is the number density of ambient (photospheric photons), $n_p$ is the number density of protons in the source, $\alpha'$ is the fine structure constant, and $Q(\delta)$ is given by 

\begin{equation}
Q(\delta)=\frac{2(11+4\delta+\delta^2)}{(1+\delta)(3+\delta)^2(5+\delta)}.
\end{equation}

\noindent For a photospheric photon density $n_\gamma\sim 10^{12}$ cm$^{-3}$ and a proton density of $n_p\sim 10^8 (10^9)$  cm$^{-3}$ ICS will exceed EIB in the 10$-$100 keV range only for very hard electron distributions: $\delta<2.9 (2.5)$. In other words, for an isotropic power-law distribution of electrons characterized by a single index $\delta$, ICS is only favored for relatively tenuous plasmas and hard electron distributions. This comparison does not include the effect of an anisotropic electron distribution. For the particular case where $\delta=3$, for favorable viewing geometries, ICS can exceed bremsstrahlung emission by a factor of a few for a pancake distribution, and 12$-$40 for a cone-beam distribution, allowing the condition on the ambient density and/or $\delta$ to be relaxed somewhat. For example, for a cone-beam anisotropy, ICS exceeds bremsstrahlung over all or part of the 10$-$100 keV range if $\delta<3.5 (3)$ for ambient plasma densities of $10^8 (10^9)$ cm$^{-3}$. Alternatively, for $\delta=3$, ICS could still contribute to the 10$-$100 keV range for an ambient density as high as $3\times 10^9$ cm$^{-3}$. It is worth pointing out that if collisions are the primary electron pitch angle scattering mechanism, low energy electrons may be nearly isotropic whereas high energy electrons could be highly anisotropic, yielding even greater enhancements of ICS relative to EIB than those noted here.  

There is no reason to suspect, moreover, that the electron distribution function is necessarily characterized by a single power law over many orders of magnitude in energy. A variety of spectral features in the electron distribution have been reported, including low- and/or high-energy cutoffs \citep{2003ApJ...586..606H}, upward breaks (a flatter spectrum) above $\sim 300-$400 keV \citep[e.g., ][]{1988SoPh..118...49D, 1988SoPh..118...95V, 1991max..conf...68R, 1998A&A...334.1099T, Ackermann2012}, and downward breaks (a steeper spectrum) above energies from $\sim 30-$200 keV \citep[e.g., ][]{1987ApJ...312..462L, 1992ApJ...389..756D, 2003ApJ...595L..97H, 2007ApJ...663L.109K}. We briefly explore conditions under which spectral breaks may be favorable to ICS emission. To do so we assume a double power-law electron energy spectrum as the input to compare the contribution from ICS and bremsstrahlung emission. For the isotropic case, the electron distribution can be written as 
\begin{equation}
f_e(\gamma, \Omega_e)=\frac{1}{4\pi}
\begin{cases}
K_L (\gamma-1)^{\delta_L}, & \gamma_1 \leqslant \gamma < \gamma_b \\
K_U (\gamma-1)^{\delta_U}, & \gamma_b \leqslant \gamma < \gamma_2 \\
\end{cases}
\end{equation} 
\noindent and $f_e(\gamma,\Omega_e)=0$ for $\gamma<\gamma_1$ and $\gamma>\gamma_2$, where $\gamma_1$ and $\gamma_2$ correspond to the lower- and upper-cutoff energies $E_1$ and $E_2$, taken to be 10 keV and 100 MeV, respectively; $K_L$, $K_U$, $\delta_L$, and $\delta_U$ are respectively the normalization constants and spectral indices in the lower and upper energy range, separated by a break energy $E_b$ (corresponding to $\gamma_b$). We first consider the case where the spectrum ``breaks down'' abve $E_b$; i.e., $\delta_L<\delta_U$. We fix the upper spectral index $\delta_U$ and vary $E_b$ and $\delta_B$ as free parameters to explore the relative contributions of ICS and thin-target, non-thermal bremsstrahlung to the HXR spectrum. In effect, the electrons below $E_b$ largely determine the bremsstrahlung contribution whereas those well above $E_b$ determine the ICS emission. Smaller values of $\delta_L$ and/or larger values of $E_b$ reduce the bremsstrahlung contribution relative to the ICS contribution.

\begin{figure}[ht]
\begin{center}
 \includegraphics[width=0.45\textwidth]{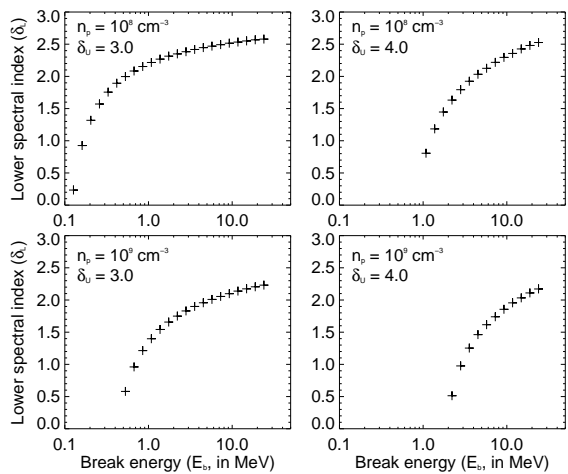}
  \caption{Comparison of ICS and bremsstrahlung emission from the same double power-law spectrum that breaks down (steepens) at higher energies. Combinations of the break energy $E_b$ and lower spectral index $\delta_L$ for a given upper spectral index $\delta_U$ are shown to equalize the integrated ICS and bremsstrahlung photon energy from 10 keV to 100 keV. The ICS wins over the bremsstrahlung emission in the lower-right region (larger $E_b$ and smaller $\delta_L$) and vice versa.}\label{fig:aic_abr_comp}
\end{center}
\end{figure}

To illustrate this we return to ICS from photospheric photons with a mono-energetic energy 2 eV and a number density of $10^{12}$ cm$^{-3}$ scattering on relativistic electrons. The ambient ion density is assumed to be $10^8$ cm$^{-3}$ (upper panels) and $10^9$ cm$^{-3}$ (lower panels). A combination of $\delta_L$ and $E_b$ is found for each given $\delta_U$ by equalizing the integrated ICS and bremsstrahlung photon emissivities from 10 keV to 100 keV. In Fig. \ref{fig:aic_abr_comp} it is seen that ICS dominates over thin-target bremsstrahlung emission in the lower-right region (larger $E_b$ and smaller $\delta_L$) and thin-target bremsstrahlung dominates in the upper left region. We also notice that for a softer spectrum at high energies (a larger $\delta_U$), the condition of equality moves to the lower-right corner of the 2-D parameter space, which means a larger $E_b$ or a smaller $\delta_L$ is required. A large value of $E_b$ increases the relative number density of higher energy electrons, increasing the contribution of ICS emission; whereas a lower value of $\delta_L$ reduces the relative number of low energy electrons leading to a corresponding reduction in bremsstrahlung emission. 

As noted above, in some cases the electron energy distribution is better characterized by a broken power law where $\delta_U<\delta_L$; that is, the spectrum ``breaks up" above $E_b$. In these cases, the electron spectrum hardens above energies of a few $\times 100$ keV and then rolls over to a steeper distribution at energies of several MeV to several tens of MeV \citep[e.g., ][]{Ackermann2012}. If the flat spectral component indeed extends to 10s of MeV ICS can quite easily exceed bremsstrahlung emission in the photon energy range of a few $\times 100$ keV and above; i.e., for continuum $\gamma$-ray emission. Even for an ambient density of $10^{10}$ cm$^{-3}$, ICS exceeds bremsstrahlung for $\delta<2.5$. 

For the case of HXR photons resulting from EUV/SXR photons up-scattered by mildly relativistic electrons, it is more difficult to assess the relative contributions of ICS and bremsstrahlung emission. It is no longer possible to treat the photon density distribution as mono-energetic, an approximation that is justified only when $\epsilon_2\gg\epsilon_1$. It is instead necessary to integrate over the photon energy distribution which, in turn, depends on the nature of the flare in question.  Consideration of ICS in the mildly relativistic regime must therefore be assessed on a case by case basis. We briefly consider two specific cases in the next section but detailed comparisons between the two mechanisms in the mildly relativistic regime are beyond the scope of this paper. 

\section{Application to Coronal HXR Sources}

In this section we briefly consider selected observations. First, however, we consolidate and summarize our results:

\begin{enumerate}

\item{In \S2.1 we computed HXR spectra resulting from ICS of photospheric photons  on an isotropic distribution of ultra-relativistic electrons. A power-law distribution of electrons with an index $\delta$ results in the well-known result of a power-law spectrum of up-scattered photons with an index $\alpha=(\delta+1)/2$. We find that the ICS photon flux increases with the heliographic longitude $\lambda$ of the source. If the electron spectrum has a high-energy cutoff, the photon spectrum has a corresponding high-energy cutoff energy. The high energy cutoff in the photon spectrum also varies with $\lambda$: the cutoff increases with increasing longitude to a maximum of $\approx 4\gamma_{max}^2 \epsilon_1$. For a source at a fixed longitude, the high energy cutoff in the photon spectrum has no dependence on $\delta$.}

\item{In \S2.2 we computed HXR spectra resulting from ICS of isotropic EUV/SXR photons on an isotropic distribution of mildly relativistic electrons - those with energies of order a few $\times100$ keV to a few MeV. We find that, while the HXR spectra are qualitatively similar to those resulting from ICS on ultra-relativistic electrons, the photon spectra are significantly steeper than $(\delta+1)/2$. In particular, for 0.2 keV photons incident on isotropic electrons with $\delta=2,3,5$, we find that the spectral indices of the up-scattered photons are 1.6, 2.2 and 3.3, respectively; for 2 keV photons the photon indices are 1.9, 2.6, and 4.2, respectively. }

\item{In \S3 we considered ICS of an isotropic distribution of photons on an anisotropic distribution of electrons in the relativistic regime. The most extreme case is that of a mono-directional ($\delta$-function) beam directed along the line of sight, which results in a photon spectrum with an index $\alpha=(\delta-1)/2$, harder by 1 than the HXR photon spectral index $\alpha=(\delta+1)/2$ resulting from ICS on an isotropic distribution of electrons; moreover, the photon flux is roughly 4-5 orders of magnitude greater than that resulting from ICS of isotropic photons on isotropic electrons. More realistic are cone-beam or pancake distributions of electrons with finite angular widths. These both result in photon spectra with indices similar to the case of scattering on an isotropic electron distribution. For a favorable viewing geometry, the cone beam yields a HXR photon flux that is enhanced by a factor $\sim4/\Delta\theta_b^2$ over the isotropic case whereas a pancake distribution yields an enhancement of $\sim 1/\Delta\theta_p$ for small $\Delta\theta_b$ and $\Delta\theta_p$ for a fixed electron number density.}

\item{We considered the case of isotropic photons scattering on an anisotropic distribution of mildly relativistic electrons in \S3. We again considered a mono-directional beam, a cone-beam, and a pancake anisotropy and compared them to the case of ICS on isotropic electrons. The mono-directional beam produces an ICS HXR spectrum that is again quite hard (indices of 1.2 and 1.4 for incident photons of 0.2 and 2 keV, respectively, for $\delta=3$), but not as hard as the ultra-relativistic case (photon index of 1.0 for $\delta=3$). The HXR photon flux is enhanced by 2.5$-$3.5 orders of magnitude over the isotropic case when EUV photons are scattered and by 1$-$2.5 orders of magnitude when SXR photons are scattered. HXR spectra resulting from a cone-beam show a broken power-law structure in which the spectrum transitions from a slope similar to the mono-directional beam case to a slope similar to the isotropic case and are enhanced by $\approx 2$ orders of magnitude relative to the isotropic case. HXR spectral slopes resulting from ICS on a pancake electron anisotropy in the mildly relativistic regime depart more subtly from the isotropic case and are enhanced by $\lesssim 1$ order of magnitude relative to the isotropic case.}

\item{In \S4, we review properties of thin-target electron-ion and electron-electron bremsstrahlung emission and calculate the emission from both isotropic and anisotropic electron distributions. The directivity of thin-target, nonthermal bremsstrahlung is less than that of ICS. By assuming the same kinds of anisotropic electron distributions as above, we found that the resulting EIB spectrum gets harder with increasing degrees of anisotropy, but the enhancement is less than that found for ICS emission, especially at low HXR photon energies. The EEB bremsstrahlung emission can be safely neglected for photon energies $\lesssim$ 100 keV for an isotropic electron distribution, but becomes more important for fast electrons with higher anisotropies. The EEB emission has a stronger directivity than the EIB, but is still less directive than the ICS emissivity. Including the EEB component leads to a flattening of the HXR spectrum at high photon energies. For high degrees of anisotropy, the emissivity can be nearly doubled when including the EEB contribution.}

\item{We compared the contributions of ICS and thin-target non-thermal bremsstrahlung to coronal HXR emissions (10$-$100 keV) in \S4 for photospheric photons scattered on a simple isotropic power-law distribution of electrons. We then showed that a double power-law electron distribution that breaks down (steepens) at higher energies with a sufficiently large break energy $E_b$ and/or a flat spectrum at lower energies can result in an excess of ICS emission relative to bremsstrahlung emission. 
An electron distribution that breaks up (flattens) at higher energies can yield ICS that exceeds bremsstrahlung emission rather easily for photon energies more than a few $\times 100$ keV. ICS emission resulting from EUV/SXR photons scattering on mildly relativistic electrons depends on the details of the photon spectrum and must be computed on a case by case basis and compared with the corresponding bremsstrahlung emission.}

\end{enumerate}

\noindent We now briefly discuss observations of several coronal HXR and continuum $\gamma$-ray sources and consider whether ICS may play a role in the observed emission. 

\subsection{Coronal $\gamma$-Ray Sources from Three Powerful Flares} 

Observations of coronal continuum $\gamma$-ray (200-800 keV) sources associated with three extremely powerful X-class flares were presented by \citet{2008ApJ...678L..63K}. The flares occurred on 2003 October 28 ($>$X17; disk center), 2005 January 20 (X7.1; W61), and 2005 September 7 ($>$X17; solar limb). Early in each event, footpoint emission dominated the continuum  $\gamma$-ray emission. Later, during the (exponential) decay phase of the $\gamma$-ray emission, the coronal source became more prominent than the footpoint emission.  The photon spectral index of the coronal source in each case was significantly harder ($\alpha\approx 1.5-$2) than that of the foot-point sources ($\alpha\approx 3-$4). The authors suggest the coronal sources result from non-thermal bremsstrahlung emission.  The photon spectral indices are at, or near, the theoretical limit of non-thermal bremsstrahlung emission and require a non-thermal electron distribution with a low-energy cutoff $E_c>1$ MeV in all cases \citep{2008A&A...486.1023B}; see also \S4.  Given that the electron transit time of MeV electrons is much shorter than their collisional loss time for typical coronal conditions in a flare, electron trapping is needed to produce the coronal source. In the specific example of the flare on 2005 January 20, \citet{2008ApJ...678L..63K} suggest that synchrotron losses dominate electron energy losses and the observed $\gamma$-ray emission between 200$-$800 keV can be explained by bremsstrahlung emission from electrons with a energies $>E_c=8$ MeV.  If the trap is stable, the emission becomes thick-target and the total energy in $>8$ MeV electrons is estimated to be $\sim 10^{28}$ ergs. It should be noted that while EEB emission contributes to this photon energy range ($\sim 20\%$ of the total emission for the isotropic case) the net effect is to harden the photon spectrum somewhat. However, for a highly anisotropic electron spectrum EEB could be responsible for a larger fraction of the continuum $\gamma$-ray emission for a favorable viewing geometry. 

MM10 considered the continuum $\gamma$-ray emission from the flare on 2005 January 20. They find that ICS represents a viable alternative to bremsstrahlung emission in the sense that only a modest number of ultra-relativistic electrons are needed to account for the observed continuum $\gamma$-ray source in terms of ICS. In particular, they estimate that as long as an (isotropic power-law) electron energy distribution function extends to $>100$ MeV, a total of $\sim 10^{31}$ electrons with energies $>0.5$ MeV are sufficient to account for the source. With a source volume of $\sim 5\times 10^{28}$ cm$^{-3}$ and an ambient density of $10^8$ cm$^{-3}$, a number density of just 200 cm$^{-3}$ $>0.5$ MeV electron is needed, or a fraction of only $2\times 10^{-6}$ of the ambient electrons. As noted in \S2.1, an error in the analysis of MM10 resulted in over-optimistic ICS emissivities. Our calculations lead to a revised estimate for the total number of electrons required to account for the observed $\gamma$-ray emission reported by \citet{2008ApJ...678L..63K} for the 2005 January 20 flare that is a factor $\sim 500$ larger ($10^5$ cm$^{-3}$) which, if the electron energy distribution indeed extends continuously to $\sim 100$ MeV implies a fraction of 0.1\% of the ambient electrons are accelerated to high energies. If the electrons responsible for the emission are trapped near the loop top, they will have a pancake-like anisotropy which, as shown in \S3 requires fewer electrons for a favorable viewing geometry. If this is the case, the required number of electrons can perhaps be reduced to 1$-2\times 10^4$ cm$^{-3}$ electrons, again normalized to a reference energy of $0.5$ MeV.  

As noted above, Krucker et al. find that the energy content of the electrons $>E_c=8$ MeV needed to account for the observed emission in terms of bremsstrahlung emission is $\sim\!10^{28}$ ergs in total. The implied number of energetic electrons required to account for the source in terms of bremsstrahlung emission, normalized to $0.5$ MeV in oder to compare it to ICS, is $\sim2\times 10^6$ cm$^{-3}$, which is 20$-$100 times the number of fast electrons needed for ICS, assuming a single power-law distribution with $\delta=3$. The ratio likewise holds for the energy content of fast electrons. The index of the electron distribution function must lie in the range $\delta=2-$3 to account for the continuum $\gamma$-ray emission in terms of ICS. We conclude our discussion of the continuum $\gamma$-ray events by suggesting that ICS of photospheric photons on relativistic electrons can account for the observed emission. It is energetically more favorable than bremsstrahlung by a factor of 20 (isotropic) to 100 (pancake anisotropy). 

\subsection{HXR Emission from High in the Corona}

An intriguing example of a coronal HXR source for which ICS may have played a role is that reported by \citet{2007ApJ...669L..49K}. A powerful flare occurred in AR 10180 when it was $40^\circ$ behind the limb on 2002 October 27, as viewed from Earth. The flare was observed directly from Mars by the Gamma-Ray Spectrometer on board the {\sl Mars Odyssey} mission. On this basis it was estimated to be comparable in SXR class to the flares on 2003 October 28 and 2005 January 20 discussed in the previous section.  A large, diffuse HXR source was observed by RHESSI. Both the footpoints and the thermal flare loops were occulted for this event - in fact, the source was high in the corona: the radial occultation height is $300''\pm 65''$ and the source centroid was $\approx 80''$ above limb, implying a source height of $\sim 0.3-$0.4 R$_\odot$. The HXR spectrum was observed up to 60 keV. The photon spectral index was relatively hard, changing systematically from 3.5$-$3 during the exponential decay phase of the event, implying an electron spectral index $\delta=2.5-$3. Based on Fig.~3 in \citet{2007ApJ...669L..49K}, in which the HXR spectrum is presented for a time during the decay phase, we estimate the photon flux at 30 keV to be 0.1 photons s$^{-1}$ cm$^{-2}$ keV$^{-1}$.  When interpreted in terms of non-thermal bremsstrahlung emission, the authors estimate that the instantaneous number density of electrons $>10$ keV required $\sim 10^7$ cm$^{-3}$, or $\sim10$\% of the total electron number density for an ambient plasma density of $\sim 10^8$ cm$^{-3}$. The lower limit to the total energy in non-thermal electrons is $2\times 10^{29}$ ergs. 

Is ICS a viable alternative for this event? For the case of ultra-relativistic electrons the photon spectral index of $\alpha= 3.1$ implies an electron spectral index is $\delta=5.2$. When normalized to 0.5 MeV, a large number of energetic electrons is required, $\sim 3\times 10^7$ cm$^{-3}$, containing $10^{31}$ ergs. We note, however, that the electrons involved in ICS in this case have energies $\gtrsim 20$ MeV. The minimum required number density of electrons with energies $>20$ MeV is far more modest, $n_{min}\sim 10$ cm$^{-3}$, containing $\sim2\times10^{26}$ ergs. An anisotropic electron distribution brings these estimates down yet further. Of course, ICS depends entirely on the details of the electron distribution function. If the electron distribution is described by a single power law ($\delta=5.2$) that extends below $\sim 2$ MeV, then it becomes energetically unfavorable compared to bremsstrahlung. In the context of a double power law with $\delta_U=5.2$, the break energy - or low energy cutoff - would need to be a few MeV in order for ICS to be relevant. Consideration of an anisotropic electron distribution can bring the cutoff down to $E_b\lesssim 500$ keV. 

\begin{figure*}[ht]
\begin{center}
\includegraphics[width=0.9\textwidth]{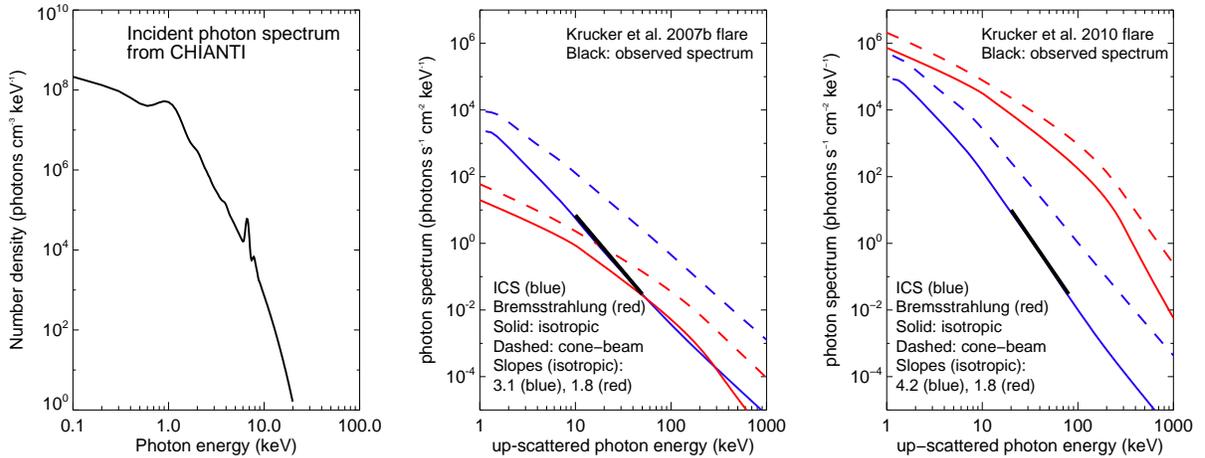} 
\caption{Left: the simulated EUV/SXR spectrum of a flare from CHIANTI (based on an M2 flare). Middle: HXR photon spectra as observed at 1 AU from the ICS (blue) and EIB (red) emission respectively, to account for the \citet{2007ApJ...669L..49K} flare. The photon spectrum has been scaled up to represent an X10 flare. The electron energy spectrum is assumed to be a double power-law extending from 10 keV to 100 MeV. The spectrum has a break energy of 300 keV, with spectral indices of 1.5 and 3.8 at the lower and higher energies, respectively. Solid and dashed lines are for an isotropic electron distribution and a cone-beam distribution with a half-width of $\theta_b=10^{\circ}$. The heavy black line represents the observed HXR spectrum. Right: HXR photon spectra from the ICS (blue) and EIB (red) emission, to account for the \citet{2010ApJ...714.1108K} flare. The electron energy spectrum has parameters similar to the middle panel, except the spectral index at higher energies ($>$ 300 keV) is now $\delta_U=5.2$. No scaling is applied to the simulated incident photon number density since the flare was itself an M2 class flare.}  \label{fig:mrel_chianti}
\end{center}
\end{figure*}

We now consider an alternative to ICS in the relativistic regime and instead consider a mildly relativistic distribution of electrons. In \S2 we showed that the up-scattered photon spectrum resulting from mildly relativistic electrons is significantly steeper than that from the classic relation $(\delta+1)/2$ at small values of $\epsilon_2/\epsilon_1$. In the mildly relativistic regime, however, mono-energetic photons are no longer appropriate as an approximation for the incident photon spectrum. Therefore, as a crude estimate, we use the simulated flare spectrum produced by the CHIANTI software package \citep{1997A&AS..125..149D,2009A&A...498..915D} based on an M2 flare in the EUV/SXR energy range (0.1 to 20 keV, see the left panel of Fig.~\ref{fig:mrel_chianti}) and scale the photon number density up to represent a large X-class flare. The EUV/SXR spectrum declines rapidly with energy above $1-2$ keV and so the redistribution of photons by down-scatter can be neglected for the HXR energy range in question. After some experimentation, we find that a double power-law electron energy distribution with an upper spectral index of $\delta_U=3.8$ produces an up-scattered HXR spectrum with the observed photon spectral index of $\alpha=3.1$ (Fig.~\ref{fig:mrel_chianti}, blue solid curve). The electron distribution is assumed to be isotropic at all energies. The spectrum has a break energy at 300 keV with a flatter spectral index at lower energies ($\delta_L=1.5$). An ambient density of $10^8$ cm$^{-3}$ has been assumed and the number density of energetic electrons with $\delta_U=3.8$ is $3\times 10^3$ cm$^{-3}$, normalized to 0.5 MeV, containing $\approx 4\times 10^{27}$ ergs - a factor $\approx 50$ less than that implied by the bremsstrahlung interpretation. The electron distribution yields EIB bremsstrahlung shown as the red solid curve in Fig.~\ref{fig:mrel_chianti}. The EIB photon spectrum is significantly flatter than the ICS spectrum, and becomes comparable to ICS at photon energies of $\gtrsim 60$ keV. The dashed lines show the ICS and EIB emission from a cone-beam (half width of $10^\circ$) distribution of electrons.  In this case, ICS clearly exceeds EIB by a large factor and the estimates of the required number density of fast electrons and their energy content is reduced by a corresponding factor. We acknowledge that the photon field resulting from a flaring active region is no longer well-approximated by an infinite plane (as was employed for the photospheric photon field). However, since ICS is dominated by large-angle scattering the error introduced is not large. We conclude that ICS from mildly relativistic electrons may make a significant contribution to the observed HXR emission in this case. Again, a broken power-law electron spectrum is required with a break energy of 300 keV. 

\subsection{Electron Acceleration in a ``Masuda"-like Event}

Another example of an intriguing coronal HXR source is the limb-occulted flare on 2007 December 31 \citep{2010ApJ...714.1108K}. The flare was $\approx 12^\circ$ over the limb as observed by RHESSI. However, comparing direct observations by the X-ray Spectrometer on board the Mercury MESSENGER spacecraft with GOES observations showed that it was an M2 SXR flare.  The flare shared many attributes with the  ``Masuda"  flare \citep{1994Natur.371..495M}, a flare that showed a HXR source above the thermal SXR loops.  The coronal HXR spectrum of the 2007 December 31 event is clearly non-thermal with a  photon spectral index 4.2 between 20$-$80 keV. In order for the thin-target bremsstrahlung model to account for the observed HXR flux, the number density of non-thermal electrons above 16 keV is required to be nearly the same as the ambient electron density ($\sim 2\times 10^9$ cm$^{-3}$). In other words, essentially all the electrons in the HXR source are accelerated to energies $> 16$ keV!

To account for the HXR spectrum in terms of ICS in the ultra-relativistic regime, a soft power-law index $\delta=7.4$ is required. With an observed HXR photon flux of $\approx$ 0.2 photons s$^{-1}$ cm$^{-2}$ keV$^{-1}$ at 50 keV and source volume of $\sim10^{27}$ cm$^3$ \citep[see Table 1 of][]{2010ApJ...714.1108K} we find that the number density of (isotropic) electrons needed is unacceptably large: $~3\times 10^{16}$ electrons $>0.5$ MeV. The electrons relevant to ICS in this regime are again those with energies $\gtrsim 20$ MeV, for which we find  $n_{min}\sim 2\times 10^6$ cm$^{-3}$. Even so, in the context of a double power law electron distribution, ICS become energetically unfavorable compared to EIB emission for break energy $E_b\lesssim 15$ MeV even with an anisotropic electron distribution. 

Turning to ICS in the mildly relativistic regime, we have again used the simulated M2 flare spectrum produced by CHIANTI. This time, however, it is not necessary to scale the flare up since the event in question was itself an M2 flare. Again considering a broken power law with $E_b=300$ keV and $\delta_L=1.5$, we find that an electron energy distribution with a spectral index $\delta_U=5.2$ produces an up-scattered HXR spectrum with the observed index $\alpha=4.2$ over the 20$-$80 keV range (Fig.~\ref{fig:mrel_chianti}, right panel). We have again ignored the finite size of the flaring active region but note that, unlike the case discussed in the previous subsection, the HXR source is relatively low in the corona and just 6 Mm above the SXR loops. The number density of energetic electrons above 0.5 MeV required to account for the HXR source is $\sim6\times 10^8$ cm$^{-3}$ for the isotropic case, which is  $\approx 30$\% of the ambient density. The total energy in mildly relativistic electrons, $\sim 3\times 10^{29}$ ergs, comparable to that estimated for the bremsstrahlung interpretation ($\gtrsim 10^{29}$ ergs). However, the bremsstrahlung emission produced by the model electron spectrum in this case exceeds the ICS contribution by orders of magnitude. This is a result of the much lower EUV/SXR photon number density for this flare, the higher ambient density, and the relatively soft electron spectrum above the cutoff. We conclude that ICS in the mildly relativistic regime cannot account for the coronal HXR source. 

\section{Concluding Remarks}

We have considered three specific cases of coronal HXR and continuum $\gamma$-ray sources. In the case of the continuum, flat-spectrum, $\gamma$-ray sources produced by the powerful flares described by \citet{2008ApJ...678L..63K}, we are in qualitative agreement with MM10 that ICS in the ultra-relativistic regime can account for the observed photon spectrum. In the case of the flare on 2002 October 27 described by \citet{2007ApJ...669L..49K}, the observed photon spectrum is somewhat softer. Either ultra-relativistic or mildly relativistic ICS may account for the observations, but an electron distribution that breaks down from a flat distribution to a steeper distribution is required in both cases. Finally, we considered the ``Masuda"-like flare reported by \citet{2010ApJ...714.1108K}. Here, ICS is unlikely to play a role in the relativistic case unless rather large numbers of ultra-relativistic electrons are accelerated and a high break energy is assumed. ICS in the mildly relativistic regime fails to account for the HXR emission in this case. 

We conclude that ICS may play a role in certain coronal HXR or continuum $\gamma$-ray sources. Such sources require rather special conditions for ICS to prevail over bremsstrahlung emission. In particular, one or more of the following conditions must be met: 1) the ambient plasma density is low; 2) the electron energy distribution is complex - a double power-law or similar; 3) the electron angular distribution is anisotropic; 4) the flare produces enough EUV/SXR photons for the mildly relativistic ICS to be effective. Sources in which ICS plays a significant role are likely rare. Potential cases must be analyzed in some detail. 

Such analyses would be greatly aided by the availability of imaging and spectroscopic data at both HXR/$\gamma$-ray energies and centimeter/millimeter radio wavelengths. We note that limited microwave observations are available for the event discussed in \S5.3; \citet{2010ApJ...714.1108K}, who find that the radio emission is broadly consistent with the bremsstrahlung interpretation, point out that the electrons that produce the 17 GHz emission have energies of order 1.2 MeV. These are the same mildly relativistic electrons that could be responsible for ICS, if it is relevant, in the mildly relativistic regime. It is well known that a close relationship between ICS and synchrotron emission exists for isotropic distributions of electrons and photons. The ratio of the synchrotron power $P_{synch}$ to the ICS power $P_{ICS}$ is equal to the ratio of the magnetic energy density $u_B = B^2/8\pi$ and the photon energy density $u_{ph}$:

\begin{equation}
\frac{P_{synch}}{P_{ICS}}=\frac{u_B}{u_{ph}}
\end{equation}

One could in principle use joint radio and HXR observations to determine whether ICS in the mildly relativistic regime is relevant. Given that anisotropic electron distributions may be an important element in determining the relevance of ICS, the above relationship would need to be recast to the specifics of the electron anisotropy. It is beyond the scope of this paper to investigate whether the HXR and radio observations of flares can be reconciled in the framework of mildly relativistic ICS, but it is important to emphasize that the same electrons are responsible for the radio emission and the photon up-scatter in this case and both mechanisms would jointly impose strong constraints on the electron distribution.

\acknowledgements
We thank Dr. Alec MacKinnon, Dr. S\"am Krucker, and the anonymous referee for useful comments that led to improvements to the paper. The National Radio Astronomy Observatory is a facility of the National Science Foundation operated under cooperative agreement by Associated Universities, Inc. CHIANTI is a collaborative project involving George Mason University, the University of Michigan (USA) and the University of Cambridge (UK). BC acknowledges support by NSF grant AGS-1010652 to the University of Virginia. 

\bibliographystyle{apj}
\bibliography{apj-jour,coronal_hxr}

\end{document}